\documentclass[11pt,a4paper]{article}
\pdfoutput=1
\usepackage{jheppub}

\newcommand{\rem}[1]{} 
\def\C{\mathbb{C}}
\def\Z{\mathbb{Z}}

\def\R{\mathbb{R}}

\def\P{\mathbb{P}}

\def\rk{\operatorname{rk}}

\def\Hirz[#1]{\mathbbm{F}_{#1}}
\def\o[#1]{\overline{#1}}

\title{M-Theory and Orientifolds}

\author{Andreas P. Braun}

\affiliation{Department of Mathematical Sciences, Durham University, Lower Mountjoy, Stockton Rd, Durham DH1 3LE, UK}

\emailAdd{durham.ac.uk:andreas.braun}

\arxivnumber{1912.06072}

\abstract{We construct the M-Theory lifts of type IIA orientifolds based on K3-fibred Calabi-Yau threefolds with compatible involutions. Such orientifolds are shown to lift to M-Theory on twisted connected sum $G_2$ manifolds. Beautifully, the two building blocks forming the $G_2$ manifold correspond to the open and closed string sectors. As an application, we show how to use such lifts to explicitly study open string moduli. Finally, we use our analysis to construct examples of $G_2$ manifolds with different inequivalent TCS realizations. }

\begin{document}

\maketitle

\section{Introduction}

The lift of type IIB orientifolds with O7$^-$-planes to F-Theory \cite{Sen:1996vd,Gopakumar:1996mu} is by now a classic result.\footnote{See \cite{Tachikawa:2015wka,Bhardwaj:2018jgp} for the F-Theory lift of O7$^+$-planes.} For a Calabi-Yau threefold $X$ and a holomorphic involution fixing a divisor on $X$, there is an associated IIB orientifold with locally cancelled Ramond-Ramond seven-brane charge that is lifted to F-Theory on the Calabi-Yau orbifold
\begin{equation} \label{eq:iiborftheorylift}
Y = \left( X \times T^2 \right)/ \Z_2 \, .
\end{equation}
Here, the complex structure of the torus encodes the value of the IIB axiodilation modulo $SL(2,\mathbb{Z})$. Deformations of $Y$ can be studied using standard techniques from algebraic geometry, and can be mapped to open and closed string moduli \cite{Sen:1997bp}. In particular, deformations of the singularities of the orbifold $Y$ can be mapped to displacements of the D7-branes away from the O7$^-$-planes. The resulting space is no longer the quotient of a product of a Calabi-Yau threefold and a torus, but only carries an elliptic fibration with a base $B$ (that can be thought of as the quotient of $X$). The locations of D7-branes and O7$^-$-planes can then be tracked by finding the degeneration loci of the elliptic fibre. 

This not only gives a geometrization of weakly coupled type IIB orientifolds, but also allows to explore their strong coupling behaviour, such as the appearance of exceptional gauge groups. In fact, starting from F-Theory on an elliptically fibred Calabi-Yau manifold at a generic point in its complex structure moduli space, an interpretation as a weakly coupled type IIB orientifold is only possible in a limit \cite{Sen:1997bp,Clingher:2012rg} which in a certain sense is close to the orbifold locus \eqref{eq:iiborftheorylift}.\footnote{For many choices of a base $B$, such a limit does not exist for the associated Calabi-Yau manifolds \cite{Halverson:2017vde}, and the resulting F-Theory models are inherently strongly coupled \cite{Morrison:1996pp,Morrison:2012np,Morrison:2012js}.} An excellent review of the relationship between F-Theory and type IIB orientifolds is given in \cite{Denef:2008wq}.

The main motivation of the present work is to develop a similar understanding for the lift of type IIA orientifolds with O6$^-$-planes to M-Theory on $G_2$ manifolds. The defining data of such orientifolds are a Calabi-Yau threefold $X$ and an anti-holomorphic involution fixing a special Lagrangian submanifold. Configurations with locally cancelled six-brane charge are lifted to M-Theory on the $G_2$ orbifolds \cite{joyce1996I,joyce1996II,Partouche:2000uq,Kachru:2001je,Gomis:2001vk,Kaste:2001iq}
\begin{equation*}
 M = \left( X \times S^1\right)/ \,\mathbb{Z}_2 \, .
\end{equation*}

Again, deformations correspond to displacements of D-branes, but it is much harder to give a general description (see \cite{2017arXiv170709325J} for the state of the art) and map it to open and closed string moduli. While this question and the relation to super-Yang-Mills theory has been studied intensively in non-compact setups \cite{Acharya:1998pm,Acharya:2000gb,Atiyah:2000zz,Edelstein:2001pu,Cvetic:2001ya,Aganagic:2001jm,Atiyah:2001qf,Acharya:2001gy,Aganagic:2001ug,Cvetic:2001sr,Cvetic:2001kp}, a concise global description at the level of detail available in F-Theory is still missing. 

We make progress by showing how to map IIA orientifolds based on K3-fibred Calabi-Yau manifolds with compatible anti-holomorphic involutions to twisted connected (TCS) sum $G_2$ manifolds \cite{MR2024648,Corti:2012kd,MR3109862}. This allows to describe deformations of the $G_2$ orbifold $M$ as resolutions or deformations of the asymptotically cylindrical Calabi-Yau manifolds used in the TCS construction. Interestingly, the decomposition of a TCS $G_2$ manifold into two asymptotically cylindrical Calabi-Yau manifolds (times a circle) is understood as the decomposition of the associated type IIA orientifold into open and closed string sectors. All of the open string degrees of freedom are hence captured by the geometry of a Calabi-Yau threefold $X_+$, and we can give a general dictionary. This in particular confirms previous studies concerning non-abelian gauge symmetry in M-Theory compactifications on TCS $G_2$ manifolds \cite{Halverson:2015vta,Guio:2017zfn,Braun:2017uku,Braun:2018vhk}. 

The dictionary between the M-Theory geometry and the IIA open string sector is of a form equivalent to the weak coupling limit of D7-branes in F-Theory. In particular, this means that we can recover several effects familiar from the physics of D7-branes in the M-Theory description of D6-branes, such as the absence of a $U(1)$ gauge boson in cases with no disjoint brane-image-brane system \cite{Grimm:2010ez,Braun:2011zm}, the folding of Dynkin diagrams to form non-simply laced gauge groups \cite{Bershadsky:1996nh,Aspinwall:2000kf} and the reduction in the naive open string degrees of freedom in the presence of an orientifold plane \cite{Braun:2008ua,Collinucci:2008pf}. 

A second application concerns the existence of TCS realizations of $G_2$ manifolds. For a given type IIA orientifold, there is a unique M-Theory lift $M$. As our construction gives a TCS $G_2$ lift for any K3 fibration on $X$ compatible with the anti-holomorphic involution $\sigma$, our methods can be used to show the existence of multiple TCS realizations of a $G_2$ manifold.

Note that we are ignoring the possibility of a membrane instanton generated superpotential \cite{Becker:1995kb,Harvey:1999as}. Such a potential can not only obstruct the deformations of the orientifold and the associated deformations of the $G_2$ lift, but can in principle lead to a potential barrier between the perturbative IIA limit and the large volume M-Theory limit. Our results suggest that such a barrier is absent or that the comparisons we are making are still physically meaningful.

Section \ref{sect:iiliftstoM} contains a general introduction to type IIA orientifolds and our central result on the M-Theory lifts of K3-fibred IIA orientifolds. Parallel displacements of the D6-branes away from the orientifold planes are always possible in the models we consider and correspond to resolutions of the Calabi-Yau threefold $X_+$ which forms the M-Theory lift of the open string sector. Using this, we are able to show the equivalence of light degrees of freedom in the IIA orientifolds and their M-Theory lifts in complete generality. 

We consider deformations of $X_+$, which map to more general deformations of the D6-branes, in Section \ref{sect:defd6}. We show how to extract the locations of D6-branes from the geometry of $X_+$ and discuss the M-Theory origin of several simple physical effects. 

In Section \ref{sect:multitcs}, we use our results about lifts of type IIA orientifolds to construct examples of $G_2$ manifolds with multiple TCS realizations. To our knowledge, these are the first instances of such a phenomenon.

To keep the paper reasonably self-contained and introduce notation, we have collected some key results on TCS $G_2$ manifolds, asymptotically cylindrical Calabi-Yau manifolds and their construction, as well as Nikulin involutions and Voisin-Borcea Calabi-Yau manifolds in the Appendices.

\section{Lifting IIA Orientifolds to TCS $G_2$ Manifolds}\label{sect:iiliftstoM}

In this section, we will consider M-Theory lifts of IIA orientifolds and show how to explicitly construct the resulting $G_2$ manifolds as twisted connected sums. To warm up, we begin by recalling a few basic facts about IIA orientifolds with O6-planes and their M-Theory lifts. 

\subsection{Review of IIA orientifolds and their M-Theory Lifts}

IIA orientifold compactifications to 4D $\mathcal{N}=1$ are constructed by modding out IIA string theory propagating on a Calabi-Yau threefold $X$ by the involution
\begin{equation}\label{eq:IIAorienti}
\mathcal{O} = \Omega_p \, (-1)^{F_L}\, \sigma\, ,
\end{equation}
where $\Omega_p$ is the operator of world-sheet parity, $F_L$ is the left-moving fermion number, and $\sigma$ is an anti-holomorphic involution of $X$. At low energies, this results in an effective 4D supergravity theory with $\mathcal{N} = 1$ supersymmetry.

As $\sigma$ is an involution it must act as an isometry on $X$ such that $\sigma^2 = 1$ and $\sigma$ being anti-holomorphic implies that it acts as
\begin{equation}
\sigma^*:\,\,\,
\begin{aligned}
&J &\rightarrow& \,\,-J \\
&\Omega^{3,0} &\rightarrow& \,\,\bar{\Omega}^{3,0} \\
\end{aligned} 
\end{equation}
on the K\"ahler form $J$ and the holomorphic three-form $\Omega^{3,0}$ of $X$. 

The fixed locus $L^\sigma$ of $\sigma$ is a real three-dimensional submanifold of $X$ (or empty), which furthermore is special Lagrangian \cite{joyce2000compact}, i.e.
\begin{equation}
\begin{aligned}
J|_{L^\sigma} &= 0 \\
\int_{L^\sigma} \Omega_{3,0} & = \int_{L^\sigma} \Re(\Omega^{3,0}) =  c\,\, \mbox{Vol}(L)
\end{aligned}
\end{equation}
where the volume of $L^\sigma$ is measure by the Calabi-Yau metric and $c$ is a normalization which does not depend on $L^\sigma$, but only on the location in the moduli space of $X$ \cite{Grimm:2004ua}. 

The action of $\sigma$ on $X$ gives a decomposition of the cohomology groups of $X$ into even and odd subspaces
\begin{equation}
\begin{aligned}
H^{1,1}(X) &=&  H^{1,1}_{+}(X) &\oplus H^{1,1}_{-}(X)\\
H^{3}(X) &=&  H^{3}_{+}(X) &\oplus H^{3}_{-}(X)
\end{aligned}\,\,\,,
\end{equation}
and the fact that $\sigma$ is an anti-holomorphic involution implies that $b^3_+(X) = b^3_-(X) = \tfrac12 b^3(X) = h^{2,1}(X) + 1$. 

The closed string moduli of a IIA orientifold are then given by truncating the spectrum of type IIA on a Calabi-Yau threefold:
\begin{center}
\begin{tabular}{c|c}
 & $\mathcal{N}=1$ multiplet \\
\hline
$h^{1,1}_+$ & $U(1)$ vector \\
$h^{1,1}_-$ & chiral \\
$h^{2,1} + 1$ & chiral
\end{tabular} 
\end{center}

Due to the combined action of \eqref{eq:IIAorienti} on space-time and the world-sheet, the special Lagrangian $L^\sigma$ is wrapped by an O6-plane, which we shall take to be an $O6^-$-plane throughout this paper. These objects are charged under the Ramond-Ramond 7-form, and the associated charge cancellation condition implies that there must be D6-branes wrapped on special Lagrangian cycles $L^i$ such that
\begin{equation}
\sum [L^i] = 2[L^\sigma] \, . 
\end{equation}
In particular, the six-brane RR charge may be cancelled locally by simply wrapping two D6-branes on $L^\sigma$. Denoting the number of components of $L^\sigma$ by $f$, such a configuration gives rise to a gauge group with algebra $\mathfrak{so}(4)^f = \mathfrak{su}(2)^{2f}$.

Deformations of the world-volume of D6-branes and Wilson lines form the moduli of the open string sector. In particular, each one-form on the special Lagrangian wrapped by a D6-brane give rise to one real deformation parameter and one Wilson line, which combine into a 4D $\mathcal{N}=1$ chiral multiplet \cite{Grimm:2011dx,Kerstan:2011dy}. Starting from a situation with locally cancelled D6-brane tadpole, let us assume that we can move all D6-branes off the O6-planes without any remaining intersections. Deformations of a D6-brane on $L^\sigma$ are described by harmonic one-forms, and the condition that we can move the D6-brane completely off the O6-plane translates to the existence of a nowhere-vanishing harmonic one-form on $L^\sigma$. In this case, each D6-brane is mapped to a disjoint image under $\sigma$ (in the covering space $X$), so that each D6-brane (in the quotient) gives rise to a $U(1)$ vector multiplet. This $U(1)$ vector is simply the Cartan of the $SU(2)$ gauge group which is present when the D6-branes are coincident with the O6-planes. In conclusion, the number of open string moduli is then given by 
\begin{center}
\begin{tabular}{c|c}
 & $\mathcal{N}=1$ multiplet \\
\hline
$2\, b^1(L^\sigma)$ & chiral \\
$2\, b^0(L^\sigma) = 2 f$ & $U(1)$ vector
\end{tabular} 
\end{center}

Let us now see what the above analysis looks like when lifting to M-Theory. IIA orientifolds in which the RR charge is cancelled locally have a lift to M-Theory on a $G_2$ orbifold $M$ given by \cite{Kachru:2001je} 
\begin{equation}
 M = \left( X \times S^1 \right)/ (\sigma, -1) \, ,
\end{equation}
i.e. the involution $\sigma$ is lifted to the M-Theory description, where it simultaneously acts as an involution on the M-Theory circle. As the latter gives rise to two fixed points on the $S^1$, $M$ has singularities locally modelled on $\C^2/\Z_2 \times \R^3$ along two copies of $L^\sigma$, so that we again find that the effective 4D $\mathcal{N}=1$ theory is a gauge theory with algebra $\mathfrak{su}(2)^{2f}$. 

It is possible to write the associative three-form $\Phi$ of $M$ in terms of the calibrating forms on $X$ and the one-form $dx$ on the $S^1$ as
\begin{equation}
\begin{aligned}
\Phi = & J \wedge dx + \Re(\Omega^{3,0}) \\
\ast \Phi = & \tfrac12 J\wedge J + \Im(\Omega^{3,0}) \wedge dx \, .
\end{aligned}
\end{equation}
Note that these forms are preserved by $\sigma$ and that $L^\sigma$ becomes an associative submanifold of $M$:
\begin{equation}
\int_{L^\sigma} \Phi = c \,\,\mbox{Vol}(L^\sigma)  \, .
\end{equation}

Ignoring the orbifold singularities, which is equivalent to focussing on the IIA closed string degrees of freedom, the Betti numbers of $M$ are 
\begin{equation}\label{eq:closedstringdof}
\begin{aligned}
b^2_c(M) &= h^{1,1}_+(X) \\
b^3_c(M) &= h^{1,1}_-(X) + h^{2,1}_+(X) + 1 \, .
\end{aligned}
\end{equation}
From the M-Theory point of view, there are $b^3(M)$ 4D $\mathcal{N}=1$ chiral multiplets and $b^2(M)$ vector multiplets, so that the above reproduces the counting of closed string degrees of freedom on the IIA side. 

The open string degrees of freedom are associated with the singularities of $M$ along $L^\sigma \times \{\pm 1\}$ and their resolution. As shown in \cite{2017arXiv170709325J}, a resolution $\tilde{M}$ of $M$ can be found for any nowhere-vanishing harmonic one-form $\lambda$ on $L^\sigma$ \cite{2017arXiv170709325J}, in which case we may write 
\begin{equation}\label{eq:g2resjoyce1}
\begin{aligned}
b^2(\tilde{M}) &= b^2_c(M) + b^2_o(M) \\
b^3(\tilde{M}) &= b^3_c(M) + b^3_o(M) \, ,
\end{aligned}
\end{equation}
with
\begin{equation}\label{eq:g2resjoyce2}
b^k_o(M) = 2 b^{k-2}(L^\sigma) \, .
\end{equation}
The factor of two stems from the fact that there are two fixed points on $S^1$. As the condition for resolvability matches with the condition previously imposed in the discussion from the IIA point of view, this precisely reproduces the counting of open string modes on the IIA orientifold. 

The topology of $\tilde{M}$ does not depend on which nowhere-vanishing harmonic one-form $\lambda$  is chosen for the resolution. This does not imply, however, that the resolution is necessarily unique, as there is a variant of this construction \cite{2017arXiv170709325J}, where the one-form $\lambda$ is taken as a section of a principal $\mathbb{Z}_2$ bundle $\mathcal{Z}$. In this case, the Betti numbers of the resolution depend on $\mathcal{Z}$ and are given by bundle valued cohomology groups:
\begin{equation}\label{eq:g2resjoyce3}
 b^k_o(M) = 2 b^{k-2}(L^\sigma,\mathcal{Z}) \, .
\end{equation}

\subsection{IIA Orientifolds which lift to TCS $G_2$ Manifolds} \label{sect:ahinvonk3fib}

In this section we analyze the case of Calabi-Yau threefolds $X$ with K3 fibrations and compatible anti-holomorphic involutions in some more detail. As we shall see in the next section, these are precisely the cases in which the M-Theory lift may be described as a twisted connected sum $G_2$ manifold. 

Let us assume $X$ has the structure of a K3 fibration, 
\begin{equation}
 S \hookrightarrow X \rightarrow_\pi \P^1_b
\end{equation}
and let the anti-holomorphic involution $\sigma$ act as $[z_1:z_2] \rightarrow [\bar{z}_1:\bar{z}_2]$ on the homogeneous coordinates of the $\P^1_b$ base. In other words, we are considering anti-holomorphic involutions which respect the K3 fibration. Restricting to the base, the fixed locus $L^\sigma$ of such involutions is always a circle\footnote{ 
This can be seen most easily by switching to a different set of homogeneous $z_1' = z_1 + i z_2$ and $z_2'= z_1 - i z_2$. In these coordinates $\sigma$ acts as $(z_1',z_2') \leftrightarrow (\bar{z}_2', \bar{z}_1')$, so that its action on the affine coordinate $z'=z_1'/z_2'$ is
\begin{equation}
 z' \rightarrow 1/\bar{z}'\, ,
\end{equation}
which fixes the circle $|z'|=1$. } $L^\sigma|_{\P^1_b} = S^1$, and we shall assume that the K3 fibration over this circle is trivial, i.e. $X$ restricted to $[z_1:z_2]=[\bar{z}_1:\bar{z}_2]$ is given by $S^1 \times S_0$ for a fixed K3 surface $S_0$. The circle $L|_{\P^1_b}$ cuts the base $\P^1_b$ into two halves and hence separates the discriminant locus of the K3 fibration (which consists of a number of points) into two sets. According to our assumption, the product of all monodromies associated with the degeneration points contained in each of the two sets must be trivial. We shall give a general construction of geometries of this type following \cite{Braun:2016igl} in Appendix \ref{app:buildingblocks}. 

\begin{figure}
 \begin{center}
 \includegraphics[height=7cm]{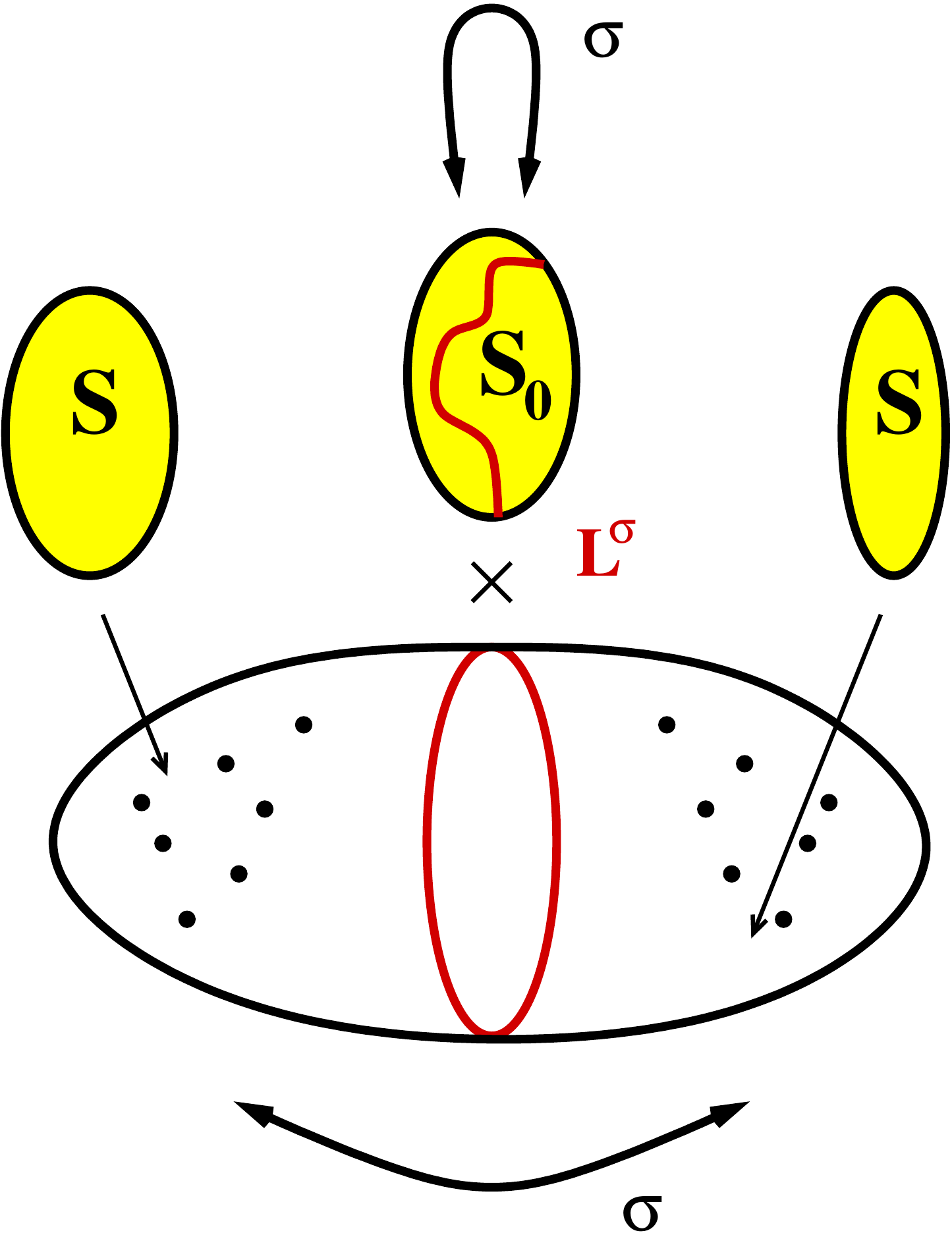}
  \caption{A cartoon of $X$ and its K3 fibration. The anti-holomorphic involution $\sigma$ identifies two hemispheres of the base $\P^1$ leaving one great circle fixed. This in particular implies that the discriminant locus of the K3 fibration is symmetric with respect to this map. Over the fixed locus of $\sigma$ in the base, $\sigma$ acts purely on the K3 fibre $S_0$ as an anti-holomorphic involution such that it keeps the special Lagrangian submanifold $L_{S_0}$ fixed point-wise. \label{fig:Xk3fib}}
\end{center}
  \end{figure}

We now describe the fixed locus $L^\sigma$ on $X$ in some detail. From the K3 fibration on $X$, it follows that $L^\sigma$ must be fibred over $L^\sigma|_{\P^1_b} = S^1$, and the assumption that this fibration is trivial implies that 
\begin{equation}
L^\sigma = L_{S_0} \times S^1 \, ,
\end{equation}
where $L_{S_0}$ is the fixed locus of $\sigma$ acting on the fibre $S_0$ over $[z_1:z_2]=[\bar{z}_1:\bar{z}_2]$. The action of $\sigma$ on $S_0$ must be an anti-holomorphic involution, i.e. it must act on the K\"ahler form $J(S_0)$ and the holomorphic two-form $\Omega^{2,0}(S_0)$ of $S_0$ as
\begin{equation}\label{eq:omegaonK3}
\sigma^*:\,\,\,
\begin{aligned}
&J(S_0) &\rightarrow& \,\,-J(S_0) \\
&\Omega^{2,0}(S_0) &\rightarrow& \,\,\bar{\Omega}^{2,0}(S_0) \\
\end{aligned} \,\,\,. 
\end{equation}
Figure \ref{fig:Xk3fib} shows a cartoon of  $X$ together with the action of the anti-holomorphic involution.
By a hyper-K\"ahler rotation, we can map \eqref{eq:omegaonK3} to an involution which is even on the K\"ahler form $J(S_0)$ and odd on $\Omega^{2,0}(S_0)$, i.e. it must be one of the non-symplectic involutions classified by Nikulin \cite{nikulin1976finite,0025-5726-14-1-A06,2004math......6536A}. According to this classification, a non-symplectic involution is characterized by a triple of integers $(r,a,\delta)$ with $r \leq 20$, $r \leq 11$ and $\delta = \{0,1\}$, see Appendix \ref{app:nikulin} for more details.  

In terms of these integers, the fixed locus of $\sigma$ on $S_0$ is given by a Riemann surface of genus
\begin{equation}\label{eq:nikg}
g =  (22-r-a)/2
\end{equation}
together with 
\begin{equation}\label{eq:nikf}
f-1 = (r-a)/2 
\end{equation}
disjoint $\P^1$s:
\begin{equation}
L_{S_0} = C_g + \sum_{i=1}^f \P^1_i \, . 
\end{equation}
The only exceptions to this rule are $(r,a,\delta)=(10,10,0)$, in which case $L_{S_0}$ is empty, and $(r,a,\delta)=(10,8,0)$, in which case $L_{S_0}$ consists of two tori. In the following, we will be mainly interested in the Betti numbers $b^1(L_{S_0})$ and $b^0(L_{S_0})$ of the fixed set. We can hence treat the latter two cases in the same language by setting
\begin{equation}
\begin{array}{ccc}
(r,a,\delta) & g & f \\
\hline 
(10,10,0) & 0 & 0 \\
(10,8,0) & 2 & 2
\end{array}\, .
\end{equation}

We can now determine the spectrum of the IIA orientifold, or, equivalently, of its $G_2$ lift $\tilde{M}$ found by resolving
\begin{equation}\label{eq:mtheoryquotient}
M = (X \times S^1)/\Z_2 \, .
\end{equation}
Using $b^0(L_{S_0}) = f$ and $b^1(L_{S_0}) = 2g$ it now follows from $L^\sigma = L_{S_0} \times S^1$ that 
\begin{equation}\label{eq:resopenstringdof}
\begin{aligned}
b^0(L^\sigma) &= f \\ 
b^1(L^\sigma) &= b^1(L_{S_0}) +  b^0(L_{S_0}) = 2g + f
\end{aligned}\, 
\end{equation}
by using the classic K\"unneth theorem. The fixed locus in \eqref{eq:mtheoryquotient} consists of two copies of $L^\sigma$, so that assuming that there exists a nowhere vanishing one-form on $L^\sigma$ we find
\begin{equation}
\begin{aligned}\label{eq:topMtilde}
b^2(\tilde{M}) &= h^{1,1}_+(X) + 2f \\
b^3(\tilde{M})+ b^2(\tilde{M}) &= 1 + h^{1,1}(X) + h^{2,1}(X) + 4g + 4f
\end{aligned}
\end{equation}
by using \eqref{eq:g2resjoyce1} and \eqref{eq:g2resjoyce2}. The  assumption we have made is implied by our earlier assumption that the K3 surface $S_0$ is not varying (metrically) over the fixed circle of $\sigma$ in the base $\P^1$ of the K3 fibration on $X$. In this case, the volume form of the fixed circle is harmonic and nowhere-vanishing on $L^\sigma$. In the language of physics, using this one-form in the resolution simply corresponds to a parallel displacement of $D6$-branes, which is the setup we considered when deriving the spectrum from the orientifold point of view. 

We will discuss other displacements of $D6$-branes in Section \ref{sect:defd6}. As these do in general not give rise to the Betti numbers \eqref{eq:topMtilde}, it follows from the results of  
\cite{2017arXiv170709325J} that these correspond $\Z_2$-twisted one-forms on $L^\sigma$. We expect that this can also be understood as discrete torsion phase for type II strings on the orbifold $M$.

\subsection{The TCS $G_2$ Lift of IIA Orientifolds }

\begin{figure}
\begin{center}
 \scalebox{.4}{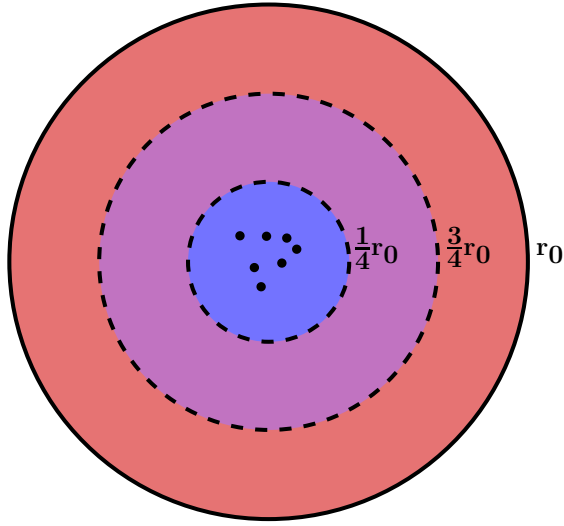}
\caption{\label{fig:D_decomp} The fundamental region of the action of $\sigma$ on $\P^1_b$ can be decomposed into two overlapping pieces $\mathbb{D}_-$ and $\mathbb{D}_+$. We are working in a limit of moduli space of $X$ where all the singular fibres of the K3 fibration are confined to a small region $r < \tfrac{1}{t} r_0$ with $t \gg 1$. Furthermore, $X$ is chosen such that the monodromy acting on the K3 fibre over the circle with $r = r_0$ is trivial.}
\end{center}
\end{figure}

We are now ready to describe $\tilde{M}$ as a TCS $G_2$ manifold and check \eqref{eq:topMtilde}. First note that $\sigma$ maps the base $\P^1_b$ of $X$ back to itself such that the fundamental region is given by a bounded disc $\mathbb{D}$, on which we can use coordinates $r,\phi$ with $r\leq r_0$ and $\phi \in \{0,2 \pi\}$. We can go into a region of the moduli space of $X$ where all of the singular fibres of the K3 fibration over $\mathbb{D}$ are contained in a small region $r \leq \tfrac{1}{t} r_0$ (with $t \gg 1$) around the origin. To find the TCS decomposition of $\tilde{M}$, we realize the disc as being glued from two open parts, see figure \ref{fig:D_decomp}. 

\begin{equation}
\begin{aligned}
\mathbb{D}_-: \{(r,\phi) | r &< \tfrac34 r_0 \} \\
\mathbb{D}_+: \{(r,\phi) | r &> \tfrac14 r_0 \} 
\end{aligned}
\end{equation}
Crucially, we can go into a limit of moduli space where the K3 fibration on $X$ becomes essentially trivial for $r > \tfrac14 r_0$. 
The decomposition of $\mathbb{D}$ now implies a decomposition of $(X \times S^1_\psi)/(\sigma,-1)$ into two parts $M_-$ and $M_+$. We claim that this decomposition is respected by a smoothing of $M$ to $\tilde{M}$ and realizes $\tilde{M}$ as a twisted connected sum. 
\begin{figure}
 \begin{center}
 \includegraphics[height=8cm]{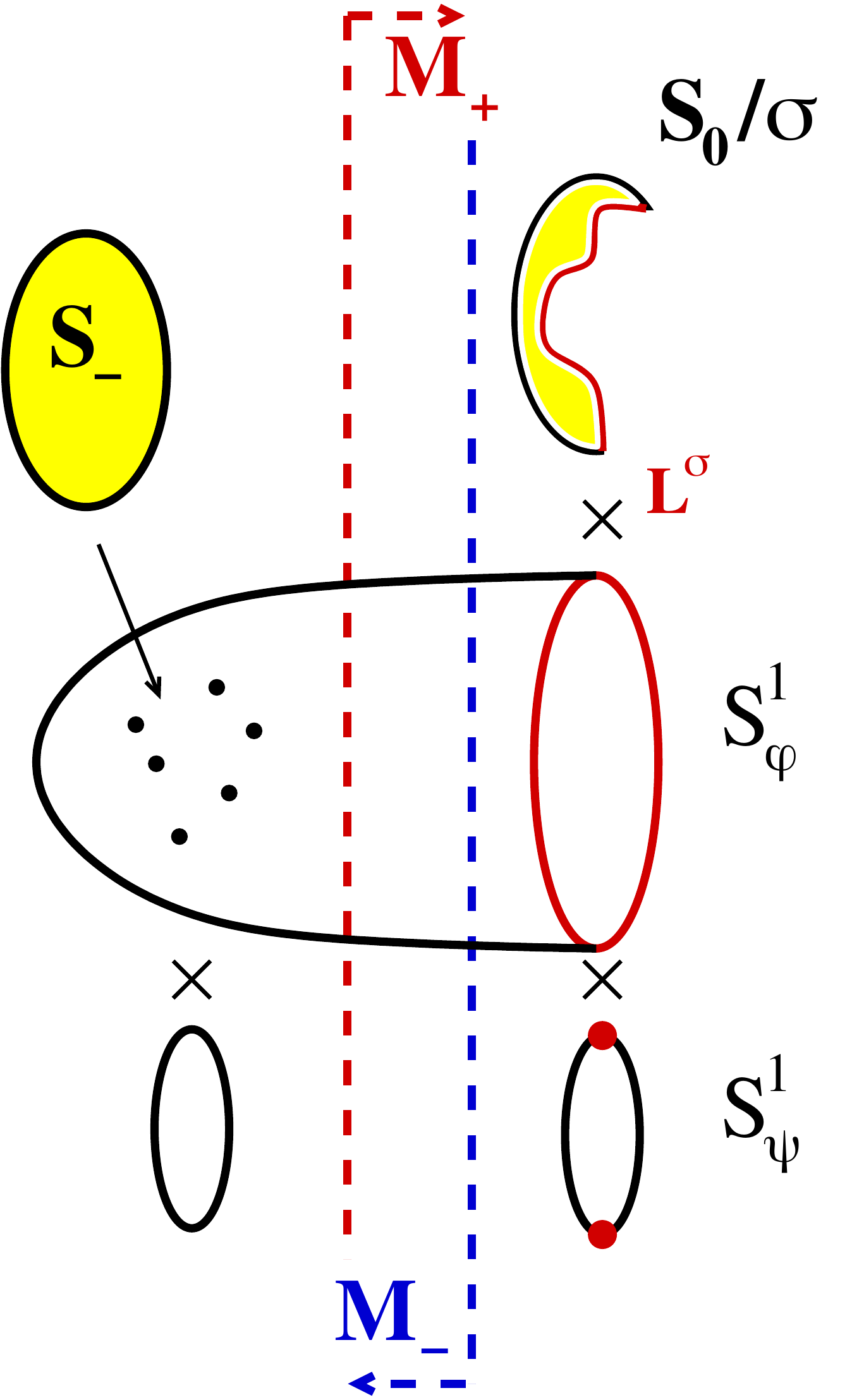}
  \caption{A cartoon of $M$, its K3 fibration, the fixed locus of $\sigma$, and its TCS decomposition into $M_-$ and $M_+$. Note that except for the action of $\sigma$, the K3 fibration is trivial on $M_+$.  \label{fig:Mk3fib}}
\end{center}
  \end{figure}

Let us first consider $M_-$. The action of $\sigma$ on the double cover of $M_-$ is free, so having restricted to a subset of a fundamental region under the action of $\sigma$ implies that we may simply write
\begin{equation}
M_- =  S^1_\psi \times   X_-\, .
\end{equation}
Here, $X_-$ is an acyl Calabi-Yau threefold which asymptotes to $S_0 \times S^1_\phi \times r$ for $r>\tfrac14 r_0 $. 

Let us now consider $\mathbb{D}_+$. The involution $\sigma$ acts as $r \rightarrow 2 r_0 - r $ and is in particular trivial on $S^1_\phi$. 
We can hence write
\begin{equation}
M_+ =  S^1_\phi \times [S_0 \times S^1_\psi \times \mathbb{R}]/\sigma  \, .
\end{equation}
The involution $\sigma$ reflects $S^1_\psi$ and $\mathbb{R}$ and acts as an anti-holomorphic involution on $S_0$. For $r < r_0$, $M_+$ is simply given by the product $S^1_\phi \times S^1 _\psi \times S_0 \times r$. Note that we may perform a hyper-K\"ahler rotation $\varphi_{r,a,\delta}$ on $S_0$ after which we obtain a K3 surface $S_0'$ with a complex structure on which $\sigma$ acts holomorphically. This hyper-K\"ahler rotation $\varphi_{r,a,\delta}$ is given by
\begin{equation} \label{eq:HK}
\begin{aligned}
\Re(\Omega^{2,0}(S_0)) & =  J(S_0') \\
\Re(\Omega^{2,0}(S_0')) & =  J(S_0) \\
\Im(\Omega^{2,0}(S_0)) & = -\Im(\Omega^{2,0}(S_0')) 
\end{aligned}\, .
\end{equation}
In the complex structure of $S_0'$, $M_+$ is given by a Calabi-Yau orbifold times $S^1_\phi$. 

The decomposition of $X \times S^1_\psi$ displayed above hence precisely realizes the structure of a TCS $G_2$ manifold as reviewed in Appendix \ref{app:TCSG2}: it can be decomposed into two halves, both of which are given by the product of an acyl Calabi-Yau threefold $X_\pm$ times a circle $S^1_\pm$ which asymptote to the products $S_{0,\pm} \times S^1_{b\pm} \times S^1_{e\pm} \times I = S_{0,\pm} \times S^1_{e\mp} \times S^1_{b\mp} \times I$. In the present setup, $S^1_{e-} = S^1_\psi$ and $S^1_{e+} = S^1_\phi$ and $S_{0,-} = S_0$, $S_{0,+} = S_0'$. Furthermore, the asymptotic K3 fibres $S_\pm$ need to be matched by precisely the hyper-K\"ahler rotation $\varphi$ in \eqref{eq:HK} we needed to apply to turn $M_+$ into the product of an acyl Calabi-Yau times a circle.

\subsubsection{Resolution of TCS and Match of Degrees of Freedom}\label{sect:resmatch}

Above, we have described how to realize the singular orbifold $M = (X \times S^1_\psi)/(\sigma,-)$ as a twisted connected sum. We are now going to describe its resolution $\tilde{M}$ in this language. As we have seen, $M_- = X_- \times S^1_{e-}$ is smooth whereas $X_+$ is singular. Such Calabi-Yau orbifolds may be resolved using standard techniques, and as long as we do not depart too far from the orbifold limit, we would expect this manifold to still have an asymptotic region in which it is described by $S_{0,+} \times S^1_{b+} \times I$. We can describe the resolution of $X_+$ (and hence a resolution of $M$) explicitly by realizing that $X_+$ is `one half' of a Voisin-Borcea Calabi-Yau threefold. Such acyl `Voisin-Borcea manifolds', i.e. resolutions of $(S \times S^1_b \times I)/\mathbb{Z}_2$, have been previously used to construct TCS $G_2$ manifolds in \cite{Kovalev_Lee}. There, it was in particular shown that resolving $(S \times S^1_b \times I)/\mathbb{Z}_2$ gives an acyl Calabi-Yau threefold. By \cite{2017arXiv170709325J} this statement is equivalent to the existence of a nowhwere-vanishing harmonic one-form on $L^\sigma$.

To find the topology of $\tilde{M}$, note that the acyl Calabi-Yau threefolds $X_\pm$ may be constructed from compact building blocks $Z_\pm$ with $c_1(Z_\pm) = [S_{0,\pm}]$ as $X_\pm = Z_\pm \setminus S_{0,\pm}$. The data we need is $h^{2,1}(Z_\pm)$, as well as
\begin{equation}\label{eq:defrhoNTK}
\begin{aligned}
\rho&: H^2(Z_\pm) \rightarrow  H^2(S_{0,\pm}) \\
N(Z_\pm) & := \mbox{im}(\rho) \\
T(Z_\pm) & := N^\perp \subset H^2(S_{0,\pm}\,\mathbb{Z})\\
K(Z_\pm) & := \ker(\rho)/[S_{0,\pm}] 
\end{aligned}
\end{equation}
and $N(Z_+) \cap N(Z_-)$, which is determined by $\varphi_{r,a,\delta}$. We will denote the ranks of the lattices $N$ and $K$ by $n$ and $k$. We have summarized details of the TCS construction in Appendix \ref{app:TCSG2}.

Voisin-Borcea Calabi-Yau threefolds \cite{voisin1993miroirs,borcea1996k3} $Y_{r,a,\delta}$ are given as a resolution of the quotient
\begin{equation}
(S \times T^2)/\eta
\end{equation}
for a K3 surface $S$ and a holomorphic involution $\eta$ which acts on $S$ by a non-symplectic involution with invariants $(r,a,\delta)$ and as $z \rightarrow -z$ on a complex coordinate on the torus $T^2$. Note that one may think of $(S \times T^2)/\eta$ as a singular version of a K3-fibred Calabi-Yau threefold over $\P^1$, the fibre of which degenerates to $S/\eta$ over $4$ fixed points. Resolving the orbifold singularities turns the fibres over these $4$ points into reducible surfaces with only normal crossing singularities. The Hodge numbers of such a resolution $Y_{r,a,\delta}$ are \cite{voisin1993miroirs,borcea1996k3}
\begin{equation}
\begin{aligned}
h^{1,1}(Y_{r,a,\delta}) & = 1 + r + 4f \\
h^{2,1}(Y_{r,a,\delta}) & = 1 + 4g + 20-r
\end{aligned}\, \,\,\, ,
\end{equation}
with $g$ and $f$ given by \eqref{eq:nikg} and \eqref{eq:nikf}. 

Voisin-Borcea threefolds may be cut into two non-compact acyl pieces $X_{r,a,\delta}$ with $X_{r,a,\delta} = \Upsilon_{r,a,\delta} \setminus S_{0}$. In terms of these `Voisin-Borcea building blocks' we may write: 
\begin{equation}
Y_{r,a,\delta} = \Upsilon_{r,a,\delta} \,\# \, \Upsilon_{r,a,\delta} \, . 
\end{equation}
This decomposition can be realized by cutting one of the two $S^1$ in the double cover in half, or equivalently, by slicing the base $\P^1$ in the quotient into two pieces, each of which contains $2$ fixed points of $\eta$. As shown in \cite{Kovalev_Lee}, $\Upsilon_{r,a,\delta}$ may also be obtained as a resolution of the quotient $S \times \P^1/\eta$. Here $\eta$ acts on the $\mathbb{P}^1$ as a holomorphic involution fixing two points. From any of the descriptions given for $\Upsilon_{r,a,\delta}$, we can find
\begin{equation}
\begin{aligned}
k(\Upsilon_{r,a,\delta}) &= 2f \\
n(\Upsilon_{r,a,\delta}) &= r \\
h^{2,1}(\Upsilon_{r,a,\delta}) &= 2g  
\end{aligned}\,\,\, .
\end{equation}

To find the data of the other building block $Z_-$, observe that one may decompose\footnote{In fact, the anti-holomorphic involution $\sigma$ swaps these two pieces. Such decompositions have been heavily exploited in \cite{Braun:2017ryx,Braun:2017csz}, they also underly the constructing of \cite{Braun:2016igl} and explain \cite{Candelas:2012uu} many of the patterns in the set of Hodge numbers in the classification of Kreuzer and Skarke \cite{Kreuzer:2000xy}.}:
\begin{equation}
X = Z_- \, \# \, Z_- \, .
\end{equation}
which implies that the topologies of $X$ and $Z_-$ are related by 
\begin{equation}\label{eq:XvsX-1}
\begin{aligned}
N(X) & = N(Z_-) \\
k(X) & = 2 k(Z_-) \\
h^{1,1}(X) &= 1 + n(Z_-) + 2 k(Z_-) \\
h^{2,1}(X) &= 2 h^{2,1}(Z_-) + 21 - n(Z_-) 
\end{aligned}\,\,\, .
\end{equation}
As the involution $\sigma$ respects the Hodge structure and the decomposition into $N$ and $T$, there is a decomposition $h^{1,1}(X) = h^{1,1}_+(X) + h^{1,1}_-(X)$, as well as 
\begin{equation}
\begin{aligned}
N(X) & \supseteq N_+(X) \oplus N_-(X) \\ 
T(X) & \supseteq T_+(X) \oplus T_-(X) 
\end{aligned}\, ,
\end{equation}
where $N(X)$ and $T(X)$ are defined in analogy to \eqref{eq:defrhoNTK}. 
Decomposing into even/odd eigenspaces one shows that the ranks of these lattices satisfy $n(X) = n_+(X) + n_-(X)$ and $t(X) = t_+(X) + t_-(X)$. The topologies of $X$ and $Z_-$ are related by
\begin{equation}\label{eq:XvsX-2}
\begin{aligned}
h^{1,1}_+(X) & = k(Z_-) + n_+(X) \\
h^{1,1}_-(X) & = k(Z_-) + n_-(X) + 1
\end{aligned}
\end{equation}
Finally, we have that 
\begin{equation}\label{eq:decompNTevenodd}
\begin{aligned}
 N(Z_+) \cap N(Z_-) & = N_+(X) \\
 N(Z_+) \cap T(Z_-) & = N_-(X) \\
 T(Z_+) \cap N(Z_-) & = T_+(X) \\
 T(Z_+) \cap T(Z_-) & = T_-(X) 
\end{aligned}
\end{equation}
This is seen as follows. Let $\eta$ be contained in $N(Z_-) \cap N(Z_+)$. This in particular implies that $\eta$ is  contained in $N(X)$.  As $\eta$ is furthermore contained in $N(Z_+)$, it must be that $\sigma: \eta \rightarrow \eta$, i.e. $\eta \in N_+(X)$. To see the converse, $\eta \in N_+(X)$ implies that $\eta \cdot \Omega^{2,0}(S_{0,\pm}) = \eta \cdot J(S_{0,\pm}) = 0$ as $J(X)$ is odd under $\sigma$ and $\eta$ is a divisor. It now follows from \eqref{eq:HK} that $\eta \in N(Z_-) \cap N(Z_+)$. The other cases can be shown by similar considerations. 

Note that \eqref{eq:decompNTevenodd} implies that the hyper-K\"ahler rotation $\varphi_{r,a,\delta}$ is such that it always corresponds to orthogonal gluing. The data determined is hence sufficient (see Appendix \ref{app:TCSG2}) to find the Betti numbers of the TCS $G_2$ manifold 
\begin{equation}
\tilde{M} =  \left[Z_- \times S^1_\psi\right] \,\, \#_{\varphi_{r,a,\delta}} \,\, \left[ \Upsilon_{r,a,\delta} \times S^1_\phi \right]
\end{equation}
as
\begin{equation}\label{eq:g2specrad}
\begin{aligned}
b^2(\tilde{M}) & = k(Z_-) + k(Z_+) + |N(Z_-) \cap N(Z_+)| = h^{1,1}_+(X) + 2f \\
b^3(\tilde{M}) + b^2(\tilde{M}) & = 23 + 2(k(Z_+) + k(Z_-)) + 2 (h^{2,1}(Z_-) + h^{2,1}(Z_+)) \\
& = h^{1,1}(X) + h^{2,1}(X) + 1 + 4f + 4g  
\end{aligned} \, .
\end{equation}
Satisfyingly, this precisely reproduces \eqref{eq:topMtilde}. Whereas the existence of a resolution leading to \eqref{eq:topMtilde} follows from the existence of a nowhere-vanishing harmonic one-form, it follows from the resolvability of Voisin-Borcea threefolds in the TCS picture. Of course, we can make contact between the two methods by noting that the nowhere-vanishing harmonic one-form needed for the resolution is simply given by $d \phi$ in the TCS\footnote{While this is certainly true for $X_+ \times S^1_\phi$, strictly speaking we need to show that the properties we want from $d\phi$ persist when we glue $M_+$ with $M_-$ to a $G_2$ manifold. }. 

Lifting a type IIA orientifold with locally cancelled D6-brane charge, like the ones considered here, the D6-branes are sitting on top of the fixed loci of anti-holomorphic involutions and hence on top of special Lagrangian submanifolds. As fixed loci of an isometric involution on $X \times S^1$, these become associative submanifolds in the M-Theory lift to $G_2$ \cite{joyce2000compact}. By a theorem of \cite{Corti:2012kd}, we can recover a similar statement for those components of the orientifold plane which are (rigid) $\P^1$s in the K3 fibre. 

\subsection{Example}

\subsubsection{The type IIA Model}\label{sect:exiia}

In order to make the previous discussion less abstract, let us consider a (reasonably simple) example. Our starting point is a Calabi-Yau threefold hypersurface $X$ which can be described as a Weierstrass elliptic fibration over $\P^1 \times \P^1$:
\begin{equation}\label{eq:ex1hseq}
X: \,\,\,y^2 = x^3 + f_{8,8}(u,z) x w^4 + g_{12,12}(u,z) w^6 \, .
\end{equation}
Here $f_{8,8}$ and $g_{12,12}$ are homogeneous polynomials of the indicated degrees in the homogeneous coordinates $[z_1:z_2]$ and $[u_1:u_2]$ of $\P_z^1 \times \P_u^1$ and $[y:x:w]$ are homogeneous coordinates on $\P^2_{123}$. In other words, $X$ is a anticanonical hypersurface in a toric variety with weight system
\begin{equation}
 \begin{array}{ccccccc}
 y & x & w & u_1 & u_2 & z_1 & z_2 \\
 \hline
 3 & 2 & 1 & 0 & 0 & 0 & 0 \\
 6 & 4 & 0 & 1 & 1 & 0 & 0 \\
 6 & 4 & 0 & 0 & 0 & 1 & 1 \\
 \end{array}
\end{equation}
and the collections of homogeneous coordinates which are forbidden from vanishing simultaneously are $(y,x,w),(u_1,u_2),(z_1,z_2)$. This toric ambient space can also be found by appropriately triangulating the reflexive polytope $\Delta^*$ with vertices
\begin{equation}
 \left(\begin{array}{rrrrrr}
-1 & 0 & 2 & 2 & 2 & 2 \\
0 & -1 & 3 & 3 & 3 & 3 \\
0 & 0 & -1 & 0 & 1 & 0 \\
0 & 0 & 0 & -1 & 0 & 1
\end{array}\right)
\end{equation}
The resulting Calabi-Yau manifold $X$ has Hodge numbers $(h^{1,1}(X),h^{2,1}(X)) = (3,243)$. It is K3 fibred over $\P^1_b$ with coordinates $[z_1:z_2]$ by a K3 surface from a lattice polarized family with polarizing lattice $U$ (the hyperbolic lattice). The divisors generating this lattice both descend from $X$, so that $N(X)=U$. The K3 fibre degenerates such that it acquires an $A_1$ singularity over 528 points. 

We can use $X$ to construct a IIA orientifold with O6-planes by specifying an anti-holomorphic involution, which we shall take to be
\begin{equation}
\sigma: (y,x,w,u_1,u_2,z_1,z_2) \rightarrow  (\bar{y},\bar{x},\bar{w},\bar{u}_1,\bar{u}_2,\bar{z}_1,\bar{z}_2) \, . 
\end{equation}
For an appropriate choice of $f_{8,8}$ and $g_{12,12}$, $\sigma$ becomes an automorphism of the hypersurface \eqref{eq:ex1hseq}. This implies that the 528 singular K3 fibres are swapped pairwise. The fixed locus of the involution consists  
of two disjoint three-tori $T^3$: on each of the two $\P^1$ factors of the base of the elliptic fibration, $\sigma$ fixes a circle. We may then choose $f_{8,8}$ and $g_{12,12}$ such that the elliptic fibration is trivial over the $T^2$ fixed locus of $\sigma$ on $\P_z^1 \times \P_u^1$, and furthermore such that $\sigma$ fixes the disjoint union of two circles in each of the elliptic fibres over the fixed locus in the base. We hence identify $L^\sigma = T^3 \bigcup T^3$, so that $b^0(L_X) = 2$ and $b^1(L_X) = 6$.  

As the involution $\sigma$ is odd on all harmonic $(1,1)$-forms of $X$, we hence find that the closed string spectrum contains 
\begin{equation}
h^{1,1}_- + h^{2,1}_+ + 1 = 3 + 243 + 1 = 247
\end{equation}
massless chiral multiplets. If we leave all D6-branes on top of the O6-planes, the open string spectrum contributes a gauge theory sector with algebra $\mathfrak{so}(4)^2 = \mathfrak{su}(2)^4$. The D6-branes can be displaced such that the gauge group is broken to $U(1)^4$ without any charged massless matter, but with twelve uncharged chiral multiplets controlling the locations of the D6-branes. We hence find the open-closed spectrum to contain 
\begin{equation}\label{eq:iiaspectrumexample}
\begin{aligned}
n_v &= 4 \\
n_c &= 259 
\end{aligned}
\end{equation}
vector, and chiral multiplets.

\subsubsection{The TCS M-Theory Lift}

Following the same arguments given in our general discussion, we can now describe the M-Theory lift of the IIA model presented above as a TCS $G_2$ manifold $M$. 

\subsubsection*{$Z_-$}

The building block $Z_-$ is given by a K3 fibration over $\P^1$ such that $c_1(Z_+) = [S_0]$. The K3 fibres are from the same lattice polarized family as the fibres of $X$, i.e. we can write
\begin{equation}
Z_-: \,\,\,y^2 = x^3 + f_{8,4}(u,z) x w^4 + g_{12,6}(u,z) w^6 \, ,
\end{equation}
as a hypersurface in an ambient toric variety with weight system
\begin{equation}
 \begin{array}{ccccccc}
 y & x & w & u_1 & u_2 & z_1 & z_2 \\
 \hline
 3 & 2 & 1 & 0 & 0 & 0 & 0 \\
 6 & 4 & 0 & 1 & 1 & 0 & 0 \\
 3 & 2 & 0 & 0 & 0 & 1 & 1 \\
 \end{array}\, .
\end{equation}
Note that this implies by adjunction that $c_1(Z_+) = [z_1]= [S_{0,-}]$. Employing the methods of \cite{Braun:2016igl}, the same hypersurface can be obtained from a projecting top with vertices
\begin{equation}
 \left(\begin{array}{rrrrr}
-1 & 0 & 2 & 2 & 2 \\
0 & -1 & 3 & 3 & 3 \\
0 & 0 & -1 & 1 & 0 \\
0 & 0 & 0 & 0 & 1
\end{array}\right)\, ,
\end{equation}
which allows to straight-forwardly compute 
\begin{equation}
\begin{aligned}
N(Z_-) & = U \\
k(Z_-) & = 0 \\
h^{2,1}(Z_-) & = 112 
\end{aligned}\,\,\, .
\end{equation}
This data can also be found by using \eqref{eq:XvsX-1} and \eqref{eq:XvsX-2} with the data of the Calabi-Yau threefold $X$ and the involution $\sigma$.

\subsubsection*{$Z_+$}

In order to describe $Z_+ = \Upsilon_{r,a,\delta}$, we first need to determine $(r,a,\delta)$. As discussed above already, $\sigma$ acts as an anti-holomorphic involution on $S_{0,-}$ with two two-tori as its fixed locus. After the hyper-K\"ahler rotation \eqref{eq:HK}, $\sigma$ will act as a non-symplectic involution on $S_{0,+}$. The fixed locus of $\sigma$ still consists of two two-tori, so that we can identify $(r,a,\delta)=(10,8,0)$. For the orientifold with locally cancelled $D6$-brane charge,$Z_+$ is hence given by the quotient $\left( S \times \P^1 \right) /  \sigma_{10,8,0}$. This building block has already been analysed, although in a different context, in \cite{Braun:2017csz,Acharya:2018nbo}. As the quotient of a K3 surface by the $(10,8,0)$ involution is a rational elliptic surface $dP_9$, it is convenient to realize $S_{0,+}$ as a double cover of $dP_9$. $Z_+$ may be then described as the complete intersection
\begin{equation}\left( S \times \P^1 \right) /  \sigma_{10,8,0}: \,\,
\begin{aligned}\label{eq:exdp6dzplus}
y^2 & = x^3 + x f_4(u) w^4 + g_6 (u) w^6 \\
\xi^2 & = u_1 u_2 z_1 z_2
\end{aligned} \,
\end{equation}
in an ambient space with weight system
\begin{equation}
 \begin{array}{cccccccc}
 y & x & w & u_1 & u_2 & z_1 & z_2 & \xi \\
 \hline
 3 & 2 & 1 & 0 & 0 & 0 & 0 & 0 \\
 3 & 2 & 0 & 1 & 1 & 0 & 0 & 1 \\
 0 & 0 & 0 & 0 & 0 & 1 & 1 & 1 \\
 \end{array}\, ,
\end{equation}
and SR ideal generated by $(u_1,u_2),(z_1,z_2),(y,x,w)$. We may choose $z_1= z_2=1$ as the asymptotic K3 surface $S_{0,+}$ used to define $X_+ = Z_+ \setminus S_{0,+}$. The monodromy upon encircling $z_1=0$ or $z_2=0$ is then precisely given by the action of $\sigma_{10,8,0}$ on $S_{0,+}$. There are four singularities of type $A_1$ located at the four two-tori sitting at $u_1 u_2= z_1 z_2 = \xi =0$. 

Resolving $\left( S \times \P^1 \right) /  \sigma_{10,8,0}$, we can find the smooth building block $\Upsilon_{10,8,0}$. Its topological data is
\begin{equation}
\begin{aligned}
k(\Upsilon_{10,8,0}) & =  4 \\
h^{2,1}(\Upsilon_{10,8,0}) & = 4 
\end{aligned}\, \,\,\,. 
\end{equation}
Resolving $\left( S \times \P^1 \right) /  \sigma_{10,8,0}$ corresponds in IIA to a parallel displacement of the $4 D6$-branes from the two $O6$-planes, breaking the gauge group to $U(1)^4$. As can be seen by explicit construction starting from \eqref{eq:exdp6dzplus}, or by realizing $S_{0,+}$ as an orbifold and following the analysis of \cite{joyce2000compact}, any smoothing of $\left( S \times \P^1 \right) /  \sigma_{10,8,0}$ will lead to a building block with the same topological data as above. This is related to the fact that the normal bundle of the fixed locus of $\sigma_{10,8,0}$ is trivial. We will describe some more interesting examples below in Section \ref{sect:defd6}. 

\subsubsection*{$\widetilde{M}$}

Having constructed both building blocks of the TCS $G_2$ manifold $\tilde{M}$, we can now work out its Betti numbers using \eqref{eq:bettitcs}. They are
\begin{equation}
\begin{aligned}
b^2(\tilde{M}) &= n_v &=& 0 + 4 + 0 &=&\,\, 4 \\
b^2(\tilde{M}) + b^3(\tilde{M}) &= n_v+n_c &=& 23 + 2(112 + 4) + 2(0+4) &=&\,\, 263
\end{aligned}
\end{equation}
Consistent with the general proof given above, these numbers reproduce the counting made from the IIA point of view in \eqref{eq:iiaspectrumexample}.

\subsection{The Weak Coupling Limit of M-Theory on TCS $G_2$ Manifolds}

The upshot of this section is that the M-Theory lift of a certain class of IIA orientifolds to M-Theory can be described as TCS $G_2$ manifolds. The only restrictions we need to put for our construction to work are that
\begin{itemize}
 \item The Calabi-Yau threefold $X$ has a fibration by K3 surfaces. The K3 fibres are from some algebraic family $\mathcal{S}$. 
 \item The anti-holomorphic involution $\sigma$ respects this K3 fibration.
\end{itemize}
Under these assumptions we may write a resolution $\tilde{M}$ of the M-Theory lift $M$ of the orientifold as 
\begin{equation}
\left[Z_- \times S^1_- \right] \,\, \#_{\varphi_{r,a,\delta}} \,\, \left[ \Upsilon_{r,a,\delta} \times S^1_+ \right]\, .
\end{equation}
Here, $Z_-$ is a building block with K3 fibres from the same algebraic family as the fibre of $X$. It is half of the Calabi-Yau threefold $X$ in the same way that a rational elliptic surface is half a K3 surface\footnote{The associated degeneration limit was constructed in appendix D of \cite{Braun:2017ryx}, see also \cite{1307.6514,Cvetic:2015uwu}.}. $\Upsilon_{r,a,\delta}$ is a `Voisin-Borcea building block'. The involution $\sigma_{r,a,\delta}$ and the hyper-K\"ahler rotation identifying the asymptotic K3 fibres are determined by the action of $\sigma$ on $X$.

The TCS decomposition of the M-Theory lift of type IIA orientifolds we have found has a beautiful physical interpretation: whereas the even cycles of $X$ under the involution $\sigma$, i.e. the closed string sector, are captured by $Z_-$, the physics of the orientifold-planes and D6-branes, i.e. the open string sector, is captured by $Z_+ = \Upsilon_{r,a,\delta}$. The Kovalev limit, in which the neck regions along which $X_\pm = Z_\pm \setminus S_{0,\pm}$ are glued grows very large, hence corresponds to a decoupling of the open and closed string degrees of freedom. The `M-Theory circle' which controls the coupling of the IIA string is given by $S^1_{e-}$. In the Kovalev limit, $S^1_{e-}$ stays at constant radius over $X_+$, so that the closed string degrees of freedom even under $\sigma$ effectively behave as in an $\mathcal{N} = 2$ Calabi-Yau compactification of type IIA strings. In contrast, the M-Theory circle becomes part of the base of the K3 fibration on $X_+$. Its radius is hence no longer constant, but becomes sensitive to the locations of O6-planes and D6-branes.

The weak coupling limit is given by shrinking $S^1_{e-}$ to small size. 
For a given TCS $G_2$ manifold, any limit in which one of the two building blocks becomes a Voisin-Borcea building block has a description as a weakly coupled type IIA orientifold. For a given $G_2$ manifold, there may be several such limits which lead to dual IIA backgrounds. From the perspective of M-Theory, such dualities simply correspond to the exchange of $Z_+$ and $Z_-$.

Note that the situation described here mirrors the weak coupling limit of F-Theory, which describes type IIB orientifolds with O7-planes and D7-branes. Such a limit was first written down by Sen \cite{Sen:1996vd,Sen:1997gv,Sen:1997bp} as a limit in the complex structure moduli space of the elliptic Calabi-Yau manifold $Y_F$ used to describe the F-Theory compactification. The limit described by Sen can furthermore be turned into a stable degeneration of $Y_F$ as shown in \cite{Clingher:2012rg}. In complete analogy to the case of M-Theory on TCS $G_2$ manifolds, the stable limit results in a decomposition of $Y_F$ into two parts, one describing the closed and one describing the open string sector.  

\section{Open String Moduli}\label{sect:defd6}

Having found the M-Theory lift of IIA orientifolds based on K3 fibrations with O6-planes and D6-branes, we now turn to discuss the lift of the open string sector to the TCS $G_2$ manifold $M$ in some more detail. As the open string sector is captured by the building block $Z_+$ in the Kovalev limit, we restrict our discussion to the geometry of $Z_+$ in this section. For an arbitrary involution $\sigma$ which acts with invariants $(r,a,\delta)\neq (10,10,0)$, the K3 surface $S_{0,+}$ can be described as a double cover over a rational surface $Y$ \cite{2004math......6536A}, branched along a smooth divisor in the class $[-2K_Y]$:
\begin{equation}
\xi^2 = \prod_{i=1}^f u_i
\end{equation}
where $u_1=0$ is a curve of genus $g$ (given by \eqref{eq:nikg}), and the remaining $f-1$ components $u_i=0$ $(i >1)$ are rational curves. The singular geometry of $Z_+$ at the orientifold point is then given by the hypersurface
\begin{equation}\label{eq:zplussingular}
\xi^2 = \left[\prod_{i=1}^f u_i \right]  z_1 z_2  \,
\end{equation}
inside an appropriate bundle (with coordinate $\xi$) over $Y \times \P^1$. The $\P^1$ with homogeneous coordinates $[z_1:z_2]$ becomes the base of the K3 fibration on $Z_+$. The example \eqref{eq:exdp6dzplus} with $(r,a,\delta)=(10,8,0)$ corresponds to the case where $Y = dP_9$.

As expected from the IIA string, the threefold $Z_+$ has two singularities of type $A_1$ correspond to the stack of an O6-plane and two D6-branes along each of the components of $[-2K_Y]$. D6-branes are objects of real codimension three in the Calabi-Yau $X$. We hence expect a general deformation to make use of all of the three transverse directions. In the previous section, we have focused on resolving the $A_1$ singularities (from the point of view of the complex structure of $Z_+$). This corresponds to a displacement of the D6-branes along one of their three transverse directions. In the picture of the TCS $G_2$ manifold, this displacement is along the interval direction. 

The acyl Calabi-Yau threefold $X_+$ is formed from $Z_+$ by excising a smooth fibre of its K3 fibration. Crucially, we cannot choose the fibres over $z_1=0$ or $z_2=0$, as these are singular. To simplify the following discussion, let us switch to the coordinates
\begin{equation}
\begin{aligned}
z_1 &= \zeta_1 + \zeta_2 \\
z_2 &= \zeta_1 - \zeta_2 
\end{aligned}
\end{equation}
In these coordinates, the singular K3 fibres of the threefold $Z_+$ are located over $\zeta_1 = \pm \zeta_2$ and we may take 
\begin{equation}
X_+ =  Z_+ \setminus \{ \zeta_2 = 0 \} \, .
\end{equation}
As $\zeta_2 \neq 0$, we can use the $\mathbb{C}^*$ action to fix $\zeta_2\ = 1$, so that $X_+$ is given by 
\begin{equation}%\label{eq:zplussingular}
\xi^2 = \left[\prod_{i=1}^f u_i \right] (\zeta_1^2-1)  \, .
\end{equation}

\subsection{Deforming $X_+$}

The singular threefold $Z_+$ may not only be resolved, but also deformed, which corresponds to a displacement of the D6-branes along its transverse directions in the K3 surface $S_{0,+}$. As long as we can make sure that $S_{0,+}$, i.e. the K3 fibre over $\zeta_2=0$ remains unchanged, we expect such deformations to lift to deformations of $M$.

As we have seen, there are $f$ disjoint O6-planes described by $u_i=0$ inside the K3 surface $S_{0,+}$. $f-1$ of these are $\P^1$s, which are rigid in K3. By Riemann-Roch and adjunction, their normal bundle has no holomorphic sections, so that there is only a single holomorphic section $u_i$ for each divisor class $[u_i]$. Let us focus on the neighborhood of one of those divisor, which we simply denote by $u$. The most general deformation can be written as
\begin{equation}
\xi^2 =  u (a\zeta_1^2 + b \zeta_1 + c)  \, ,
\end{equation}
for some constants $a,b,c$ (linear terms in $\xi$ can always be eliminated by a shift in $\xi$). This deformed equation however still has the same two $A_1$ singularities, which have merely been displaced: this can be seen explicitly by factoring the quadratic polynomial in $\zeta_1$, which is always possible as $a,b,c$ are constants. Hence any of the $f-1$ $\P^1$ components in the O6-plane locus can only be resolved in the M-Theory lift, but not deformed. Note that the O6-plane locus in such a deformation is still at $u=0$. To keep the asymptotic K3 surface at $\zeta_2=0$ fixed, we need to demand $a=1$. 

Let us hence focus on the component of the orientifold plane which is a curve of genus $g$ with $g\geq 1$. In this case, we may write a general deformation as 
\begin{equation}\label{eq:defx+}
\xi^2 =  \zeta_1^2 h + \zeta_1 \eta + \chi \, ,
\end{equation}
with $h,\eta,\chi$ different sections of the bundle $[u]$. For a generic choice, this deforms both of the two $A_1$ singularities which are present at the orientifold point, so that $Z_+$ and hence $M$ becomes smooth.

\subsection{Deformations of $X_+$ as D6 Moduli}

We expect a deformation of $M$ inherited from a deformation of $Z_+$ (as opposed to a resolution) to move the D6-branes away from the O6-plane in the direction of the K3 fibre of $X_+$. To understand in some more detail where they have been moved, we need to keep track of the degenerations of the M-theory circle, $S^1_-$. This circle is identified as the $S^1$ in the $[z_1:z_2]$ plane which degenerates to $S^1/\Z_2$ over the locations $u_i=0$ in \eqref{eq:zplussingular}. Once we start deforming \eqref{eq:defx+}, we can still identify a circle of minimal volume for any given $u$ which becomes $S^1_-$ far away from the origin. In fact, \eqref{eq:defx+} is just a deformed singularity of type $A_N$ (with $N$ depending on the degrees of $h,\eta,\chi$). Such a singularity describes the M-theory lift of D6-branes located at 
\begin{equation}\label{eq:d6loci}
\Delta \equiv \eta^2 - 4 h \chi = 0 \, .
\end{equation}
Said differently, \eqref{eq:defx+} describes a double cover over the $\zeta_1$-plane branched over two points for every value of $u$ (i.e.  $h,\eta,\chi$). This space contains a circle of minimal radius which measures the distance between the two branch points. Identifying this circle with the M-Theory circle, D6-branes are located where the two branch points meet, which happens whenever $\Delta = 0$. 

We are now ready to give the dictionary between the open string sector of a IIA orientifold and its TCS $G_2$ M-Theory lift. The building block $Z_+$ describes the open string sector, and takes the general form
\begin{equation}
%\label{eq:defx+}
\xi^2 =  \zeta_1^2 h + \zeta_1 \zeta_2 \eta + \zeta_2^2 \chi \, , 
\end{equation}
as a double cover over $Y \times \mathbb{P}^1$. Here, $h,\eta,\chi$ are sections of $[-2K_Y]$, and the acyl Calabi-Yau threefold $X_+$ with asymptotic fibre $S_{0,+}$ is found by excising the K3 fibre over $\zeta_2 = 0$. This implies that the asymptotic fibre $S_{0,+}$ is a double cover over $Y$, branched at $h=0$, i.e. the location of the O6-planes on the $u$-plane is $h=0$ and the asymptotic K3 fibre $S_{0,+}$ stays fixed as long as we keep $h$, i.e. the O6-plane, fixed. The locations of the D6-branes in the $u$-plane are given by $\Delta = \eta^2 - 4 h \chi = 0$. 

Note that these formulae are identical to the description of D7-branes and O7-planes which appears in the weak coupling limit of F-Theory \cite{Sen:1997bp,Clingher:2012rg}. This comes as no surprise, as we may convert IIA orientifolds into IIB orientifolds by mirror symmetry, which can then be lifted to F-Theory. Note that the particular form of \eqref{eq:d6loci} forces D-branes to be tangential at their intersections with O-planes \cite{Braun:2008ua,Collinucci:2008pf}. It is possible to argue for this behavior by consistency of probe branes or by carefully constructing the sheaf describing the D-branes on the orientifold \cite{Collinucci:2008pf}. We expect similar methods to apply in the present case.

The geometrisation of D6-branes in terms of the $G_2$ manifold $M$ allows us to match the deformations of $Z_+$ with deformations of the D6-brane locus. We can determine the number of D6-brane moduli and the number of open string $U(1)$s from the topology of $Z_+$ as follows. Using \eqref{eq:XvsX-1} and \eqref{eq:XvsX-2}, we may express the topology of $M$ (we continue to denote a deformation of the orbifold by $M$ in order not to clutter notation) as 
\begin{equation}
\begin{aligned}
 b^2(M) & = h^{1,1}_+(X) + k(Z_+) \\ 
 b^2(M) + b^3(M) & = 1+ h^{1,1}(X) + h^{2,1}(X) + 2h^{2,1}(Z_+) + 2 k(Z_+)
\end{aligned}\,\, .
\end{equation}
Comparing with \eqref{eq:closedstringdof}, we can hence make the following identifications between the open string sector and the topology of $Z_+$:
\begin{equation} \label{eq:openstringmodulifromZ+}
\begin{aligned}
b^2_o(M) & = k(Z_+) \\    
b^2_o(M) + b^3_o(M) & = 2 k(Z_+) + 2 h^{2,1}(Z_+)\\    
\end{aligned}\, .
\end{equation}
Here $b^2_o(M)$ counts the number of open string $U(1)$s and $b^3_o(M)$ counts the number of open string moduli.

As we have seen, the fixed locus of the orientifold involution on $S_{0,+}$ is in general reducible into several $\P^1s$ and a curve $C_g$ of genus $g$. As only D6-branes wrapped on $C_g$ may be deformed, we will limit our discussion to them in the following. At the orientifold point, there are two D6-branes (in the quotient picture) wrapped on the fixed locus of the involution and $Z_+$ is simply an orbifold of $K3 \times \P^1$. This is realized by making the choice $\eta = \chi = h$. 

In contrast, the `generic' case is to make the most general choice of $\eta$ and $\chi$. The two D6-branes wrapped on $h=0$ then become recombined into a single object described by \eqref{eq:d6loci}. Note that this implies that the D6-brane in the double cover does not split into a brane and an image brane, so that the analysis of \cite{Grimm:2011dx,Kerstan:2011dy} does not apply. For a D6-brane which is not split into a disjoint brane and its image, the $U(1)$ gauge vector is projected out by the orientifold involution. Furthermore, as the D6-brane is described by \eqref{eq:d6loci} instead of a generic divisor in its class $2[u]$, we expect that its moduli are no longer counted by $b^1(L)$, but will be constrained to a smaller subset. As we will demonstrate in examples below, these expectations are indeed met. Although we do not pursue this here, it should be possibe to describe this in the language of \cite{2017arXiv170709325J} as using an appropriately $\Z_2$-twisted one-form on $L$ to resolve $M$.

There are many intermediate cases with abelian or non-abelian gauge enhancement that can be engineered by choosing appropriate polynomials $\eta$ and $\chi$. For weakly coupled IIA strings, we expect the only gauge algebras that can arise to be $\mathfrak{su}(n)$, $\mathfrak{so}(m)$, and $\mathfrak{sp}(k)$. This is reflected in the fact that the only ADE types that occur in \eqref{eq:d6loci}, and that are not entirely contained the K3 fibres\footnote{Such singularities correspond to closed string gauge enhancement.} are $A_N$ and $D_N$, but never $E_6,E_7,E_8$. Similar to the F-Theory setting \cite{Bershadsky:1996nh}, these can be `split' or `non-split' which allows to further realize the gauge groups $Sp(n)$, and $SO(n)$ for $n$ odd. The mechanism is the same as in F-Theory: there is a monodromy action on the exceptional $\P^1$s of the resolution which can be translated to an outer automorphism  which folds the associated Dynkin diagram. Some example of this type in the TCS $G_2$ context are contained in \cite{Braun:2017uku}.

\subsection{An Example}

Let us work through an example to bring the discussion of the last section to life. As we will only be interested in the geometrisation of the open string sector, we continue to limit the discussion to $Z_+$ (i.e. we keep the O6-plane fixed). Consider the involution with invariants $(r,a,\delta) = (1,1,0)$. The fixed locus of $\sigma_{1,1,0}$ on $S_{0,+}$ is given by a curve of genus $10$ and we may realize $S_{0,+}$ as a double cover of $Y = \P^3$. $Z_+$ at the orientifold point is
\begin{equation}\label{eq:exd6modorientifoldpt}
\xi^2 = z_1 z_2 P_6(u_i) 
\end{equation}
as a hypersurface in an ambient space with weight system 
\begin{equation}
 \begin{array}{cccccc}
 u_1 & u_2 & u_3 & \xi & z_1 & z_2\\
 \hline
 1 & 1 & 1 & 3 & 0 & 0  \\
 0 & 0 & 0 & 1 & 1 & 1
 \end{array}\, 
\end{equation}
and SR ideal generated by $(u_1,u_2,u_3),(z_1,z_2)$.

A resolution of the two $A_1$ singularities yields the space $\Upsilon_{1,1,0}$ with 
\begin{equation}
\begin{aligned}
k(\Upsilon_{1,1,0}) & = 2\\
h^{2,1}(\Upsilon_{1,1,0}) & = 20
\end{aligned}
\end{equation}
The resolution corresponds to a parallel displacement of the D6-branes away from the O6-plane, which results in a theory with gauge group $U(1)^2$. These gauge bosons originate from the contribution of $k(\Upsilon_{1,1,0})$ to $b^2(M)$. Furthermore, there are $b^1(D6) = b^1(C_{10} \times S^1) = 21$ chiral multiplets for each of the two D6-branes. 

The open string sector hence contributes $2$ $U(1)$ vector multiplets and $42$ chiral multiplets. From \eqref{eq:openstringmodulifromZ+}, this matches the contribution of $Z_+$ to the Betti numbers of $\tilde{M}$.

Insted, let us now consider a generic deformation of \eqref{eq:exd6modorientifoldpt} to 
\begin{equation}\label{eqdefexample}
\xi^2 =  \zeta_1^2 h + \zeta_1 \zeta_2 \eta + \zeta_2^2 \chi \, , 
\end{equation}
where now $h$, $\eta$ and $\chi$ are all homogeneous polynomials of degree $6$ on $\P^2$. This completely smoothes the orbifold singularities of \eqref{eq:exd6modorientifoldpt}. The resulting building block $Z_+$ can also be constructed from a projecting top with vertices
\begin{equation}
 \left(\begin{array}{rrrrr}
-3 & -1 & 0 & 0 & 1 \\
-1 & 0 & 0 & 1 & 0 \\
-1 & 0 & 1 & 0 & 0 \\
0 & 1 & 0 & 0 & 0
\end{array}\right)\, .
\end{equation}
It follows that
\begin{equation}
\begin{aligned}
k(Z_+) &= 0 \\
h^{2,1}(Z_+) & = 54
\end{aligned}\, .
\end{equation}
The vanishing of $k(Z_+)$ implies that there are no open string $U(1)$s. This confirms our expectation that the world-volume $U(1)$ vector is projected out by the orientifold. There are $108$ chiral multiplets associated with open string moduli. In the $G_2$ manifold, half correspond to deformations of the Ricci-flat metric (which map to displacement of the D6-brane) and half map to moduli of the 3-form (which map to Wilson lines on the D6-brane). 

The D6-brane is given by a circle ($S^1_+$), times the vanishing of the degree $12$ polynomial $4 h\chi - \eta^2$ in $\P^2$. This implies that its first Betti number is $1 +  11\cdot 10 = 111$. The standard logic hence implies that there are $111$ chiral multiplets associated with deformations and Wilson lines, which overshoots the correct value of $108$. This is the IIA analogue of an observation made in \cite{Braun:2008ua,Collinucci:2008pf} in the context of D7-branes intersection O7-planes. Working out the topology of the D6-brane in the double cover and the action of the orientifold on its world-volume, one can observe however that the number of even one-cycles is given by $108$. We leave a general analysis of this phenomenon along the lines of \cite{Braun:2009bh} to future work.

Finally, it is possible to count the number $n_{poly}$ of polynomial deformations of \eqref{eq:defx+}. These are necessarily complex. To compare with the number of chiral multiplets, observe that the real number of deformation degrees of freedom needs to match the number of chiral multiplets in the open string sector, i.e. there must be twice as many chiral multiplets as complex deformations. 

The number of polynomial deformation corresponding to moving the D6-brane are given by the number degrees of freedom in the polynomials $\eta$ and $\chi$. As we keep the O-plane and $S_{0,+}$, and hence $h$, fixed, there are no redundancies from redefining coordinates on $Y$, and the only coordinate redefinition we can do is $\zeta_1 \rightarrow \zeta_1 + a \zeta_2$ for a complex number $a$ (recall that $S_{0,+}$ is defined by $\zeta_2=0$. Finally, we may rescale \eqref{eqdefexample} by a complex number. This leaves us with $2 \cdot (7\cdot 8)/2 - 2 = 54$ complex deformation degrees of freedom of the D6-brane, which must sit in $108$ chiral multiplets\footnote{Recall that each chiral multiplet contains one deformation degree of freedom together with a mode of the $C$-field.} This perfectly matches our previous analysis.

\section{$G_2$ Manifolds with Multiple TCS Decompositions}\label{sect:multitcs}

In this section we present an different application of our M-Theory lifts. To the knowledge of the author, the question if there are $G_2$ manifolds that admit several TCS decomposition has so far not been answered in the literature. As we have seen, a type IIA orientifold based on a K3-fibred Calabi-Yau threefold $X$, together with a compatible anti-holomorphic involution $\sigma$, lifts to a TCS $G_2$ orbifold $M$. Whereas the building block $Z_+$ is determined by the action of $\sigma$ on the K3 fibres of $X$, $X_+ = Z_+ \setminus S_{0,+}$ is one half of the Calabi-Yau threefold $X$. The construction of $X_+$ crucially employs the K3 fibration in that $X_+$ is obtained from $X$ by cutting along an $S^1$ in the base of the K3 fibration. 

This implies the following possibility: suppose that $X$ admits two K3 fibrations which are both compatible with $\sigma$. We can then use our construction of a TCS $G_2$ lift with respect to either of these two K3 fibrations. Although $Z_-$ will be the same in both cases, the building block $Z_+$ will be different in general. As the $G_2$ lift $M$ (or a smoothing/resolution $\tilde{M})$ is independent of any TCS decomposition, this implies that $M$ has several inequivalent TCS decompositions, one for each K3 fibration on $X$ compatible with $
\sigma$. Instead of developing a general framework, we limit ourselves to describing two examples of this phenomenon in the following.

\subsection{Example 1}

As our first example, consider a Calabi-Yau threefold $X$ from the family defined by the reflexive polytope $\Delta^\circ$ with vertices
\begin{equation}
\Delta^\circ =  \left(\begin{array}{rrrrrr}
-1 & 0 & 2 & 2 & 2 & 2 \\
0 & -1 & 3 & 3 & 3 & 3 \\
0 & 0 & -6 & 0 & 6 & 0 \\
0 & 0 & 0 & -1 & 0 & 1
\end{array}\right)\, .
\end{equation}
The associated Calabi-Yau hypersurface $X$ has Hodge numbers 
\begin{equation}
h^{1,1}(X) = h^{2,1}(X) = 43 \, .  
\end{equation}
K3 fibrations on $X$ (or rather, at least some of them) can be detected by studying reflexive sub-polytopes of $X$ \cite{Candelas:1996su,Avram:1996pj}. $\Delta^\circ$ has reflexive subpolytopes $\Delta_{F a}^{\circ}$ and $\Delta_{F b}^{\circ}$ with vertices
\begin{equation}
\Delta^{\circ}_{F a} =\left(\begin{array}{rrrr}
-1 & 0 & 2 & 2 \\
0 & -1 & 3 & 3 \\
0 & 0 & -6 & 6 \\
0 & 0 & 0 & 0
\end{array}\right) \hspace{.5cm} 
\Delta^{\circ}_{F b} =
\left(\begin{array}{rrrr}
-1 & 0 & 2 & 2 \\
0 & -1 & 3 & 3 \\
0 & 0 & 0 & 0 \\
0 & 0 & -1 & 1
\end{array}\right)\, .
\end{equation}

For appropriate triangulations, these subpolytopes define K3 fibrations on $X$ and we may define a compatible anti-holomorphic involution as follows: let us denote the homogeneous coordinate associated with a lattice point $\nu$ on $\Delta^\circ$ by $z(\nu)$. An anti-holomorphic involution which acts as on the base coordinates $[b_1:b_2]$ of both K3 fibrations as $[b_1:b_2] \rightarrow [\bar{b}_2:\bar{b}_1]$ is then given by
\begin{equation}\label{eq:ahinvexmultitcs}
\sigma: z(\nu) \rightarrow \overline{z(R \nu)} \, ,
\end{equation}
where $R$ is the matrix $R = \mbox{diag}(1,1-1,-1)$. The fixed point set of $\sigma$ is given by the union of two three-tori, i.e. $\sigma$ acts with invariants $(10,8,0)$ on both of the K3 fibres. It hence follows from \eqref{eq:topMtilde} that a resolution\footnote{For involutions with invariants $(10,8,0)$, the resolution of $Z_+$ has the same topological numbers as a deformation of $Z_+$.} $\tilde{M}$ of $M = X \times S^1 / (\sigma,-)$ has Betti numbers
\begin{equation}\label{eq:bettiex1difftcs}
\begin{aligned}
b^2(\tilde{M}) & = 24 \\
b^2(\tilde{M}) + b^3(\tilde{M}) & = 103
\end{aligned}\, .
\end{equation}

Let us now study the two different TCS decompositions of $\tilde{M}$ and reproduce \eqref{eq:bettiex1difftcs}. For both of them, the building block $Z_+$ is given by $\Upsilon_{10,8,0}$ with $k(\Upsilon_{10,8,0}) = h^{2,1}(\Upsilon_{10,8,0}) = 4$. Now consider the fibration implied by $\Delta^{\circ}_{F a}$. The lattice $N_a(X)$ and $k_a(X)$ of this K3 fibration is
\begin{equation}
N_a(X) = U \oplus (-E_8)^{\oplus 2} \,, \hspace{.5cm} k_a(X) = 24 \, . 
\end{equation}
The involution $\sigma$ acts on this fibration such that $k_\pm(X) = 12$ and $n_+(X)=8$. 
Using \eqref{eq:XvsX-2}, we can now compute 
\begin{equation}
k(Z_{-a}) = 12 \, , \hspace{.5cm} h^{2,1}(Z_{-a}) = 20
\end{equation}
as well as $|N(Z_+) \cap N(Z_-)| = n_+(X) =  8$. The data of $Z_{a-}$ can also be reproduced from a projecting top $\Diamond^\circ_a$ with vertices
\begin{equation}
\Diamond^\circ_a =  \left(\begin{array}{rrrrr}
-1 & 0 & 2 & 2 & 2 \\
0 & -1 & 3 & 3 & 3 \\
0 & 0 & -6 & 6 & 0 \\
0 & 0 & 0 & 0 & 1
\end{array}\right)
\end{equation}
This data reproduces \eqref{eq:bettiex1difftcs} using \eqref{eq:bettitcsz} and \eqref{eq:bettitcs}. 

Let us now study the fibration implied by $\Delta^{\circ}_{F b}$. The lattice $N_b(X)$ and $k_b(X)$ of this K3 fibration is found to be
\begin{equation}
N_b(X) = U  \,, \hspace{.5cm} k_b(X) = 20 \, . 
\end{equation}
The involution $\sigma$ now acts such that $k_\pm(X) = 20$ and $n_+(X)=0$. 
Using \eqref{eq:XvsX-2}, we can now compute 
\begin{equation}
k(Z_{-b}) = 20 \, , \hspace{.5cm} h^{2,1}(Z_{-b}) = 12
\end{equation}
as well as $|N(Z_+) \cap N(Z_-)| = n_+(X) =  0$. The data of $Z_{b-}$ can again be reproduced from a projecting top $\Diamond^\circ_b$ with vertices
\begin{equation}
\Diamond^\circ_b =  \left(\begin{array}{rrrrr}
-1 & 0 & 2 & 2 & 2 \\
0 & -1 & 3 & 3 & 3 \\
0 & 0 & -1 & 1 & 0 \\
0 & 0 & 0 & 0 & 6
\end{array}\right)
\end{equation}
This data again reproduces \eqref{eq:bettiex1difftcs} using \eqref{eq:bettitcsz} and \eqref{eq:bettitcs}. 

We have hence found two different building blocks $Z_{-a}$ and $Z_{-b}$ such that 
\begin{equation}
\tilde{M} = \left(Z_{-a} \times S^1_- \right) \#_{\varphi_a} \left(\Upsilon_{10,8,0} \times S^1_+ \right) 
=  
\left(Z_{-b} \times S^1_- \right) \#_{\varphi_b} \left(\Upsilon_{10,8,0} \times S^1_+ \right) \, ,
\end{equation}
i.e. $\tilde{M}$ has two different realizations as a TCS $G_2$ manifold.

\subsection{Example 2}

Let us now consider a variant of the previous example.  Consider the reflexive polytope $\Delta^\circ$ with vertices
\begin{equation}
\Delta^\circ =  \left(\begin{array}{rrrrrr}
-1 & 0 & 2 & 2 & 2 & 2 \\
0 & -1 & 3 & 3 & 3 & 3 \\
0 & 0 & -3 & 0 & 3 & 0 \\
0 & 0 & 0 & -1 & 0 & 1
\end{array}\right)\, .
\end{equation}
We may define two different K3 fibrations together with a compatible anti-holomorphic involution in the same fashion \eqref{eq:ahinvexmultitcs} as for the example above. The Hodge numbers of $X$ are
\begin{equation}
h^{1,1}(X) = 11\,, \hspace{.5cm} h^{2,1}(X) = 107 \, .
\end{equation}
As $h^{1,1}_+(X) = 4$, the Betti numbers of $\tilde{M}$ are
\begin{equation}\label{eq:bettiex2difftcs}
\begin{aligned}
b^2(\tilde{M}) & = 8 \\
b^2(\tilde{M}) + b^3(\tilde{M}) & = 135
\end{aligned}\, .
\end{equation}

Let us start by analysing the first fibration. The $E_8$ lattices in $N(X)$ are replaced by two $E_6$ lattices, and as seen from the Hodge numbers of $X$, there are monodromies acting on the $E_6$ roots as the outer automorphism whose quotient is $F_4$. We hence find that $n(X) = 10$ and $k(X) =0$. As the anti-holomorphic involution swaps the two $E_6$s, we furthermore find that $n_+(X) =4$. From this, or equivalently from a top with vertices
\begin{equation}
\Diamond^\circ_{a} = \left(\begin{array}{rrrrr}
-1 & 0 & 2 & 2 & 2 \\
0 & -1 & 3 & 3 & 3 \\
0 & 0 & -3 & 3 & 0 \\
0 & 0 & 0 & 0 & 1
\end{array}\right)\, , 
\end{equation}
one finds that 
\begin{equation}
k(Z_{-a}) = 0 \, , \hspace{.5cm} h^{2,1}(Z_{-a}) = 48 \, .
\end{equation}
This reproduces \eqref{eq:bettiex2difftcs}.

The second K3 fibration is such that $N(X) = U$ and $N_+(X) = 0$, so that $k(X) = 8$. From this or from a top with vertices
\begin{equation}
\Diamond^\circ_{b} = 
\left(\begin{array}{rrrrr}
-1 & 0 & 2 & 2 & 2 \\
0 & -1 & 3 & 3 & 3 \\
0 & 0 & -1 & 1 & 0 \\
0 & 0 & 0 & 0 & 3
\end{array}\right)
\, , 
\end{equation}
one finds that 
\begin{equation}
k(Z_{-b}) = 4 \, , \hspace{.5cm} h^{2,1}(Z_{-b}) = 44 \, .
\end{equation}
This again reproduces \eqref{eq:bettiex2difftcs}.

We have found a second example of a $G_2$ manifold that allows two different TCS realizations
\begin{equation}
\tilde{M} = \left(Z_{-a} \times S^1_- \right) \#_{\varphi_a} \left(\Upsilon_{10,8,0} \times S^1_+ \right) 
=  
\left(Z_{-b} \times S^1_- \right) \#_{\varphi_b} \left(\Upsilon_{10,8,0} \times S^1_+ \right) \, .
\end{equation}

\section{Discussion and Future Directions}

In this paper, we have uncovered the relationship between a large class of type IIA orientifolds and compact TCS $G_2$ manifolds. The two pieces from which the TCS $G_2$ manifold is glued beautifully correspond to the open and closed string sectors. Correspondingly, the M-Theory circle appears as a product on the closed string side, but has a non-trivial behavior for the piece describing the open string sector. 

We found perfect agreement between the massless spectra in situations where the D6-branes are displaced from the O6-planes in parallel. For more general deformations, we found that the D6-branes have an equivalent description, and hence are subject to the same constraints, as D7-branes in IIB orientifolds \cite{Braun:2008ua,Collinucci:2008pf}. This should come as no surprise: the part of the M-Theory geometry responsible for describing the open string degrees of freedom is an orbifold of the product of a K3 surface, an interval, and a torus. We may hence T-dualise to IIB after reducing M-Theory to IIA.

Our results show how M-Theory on TCS $G_2$ manifolds is linked to weakly coupled IIA orientifolds (possibly after singular transitions). Contrary to F-Theory, it seems much harder to uncover the emergence of exceptional gauge groups in strongly coupled situations in the present case. In the F-Theory description of IIB orientifolds, a crucial observation leading to the engineering of exceptional gauge groups is the non-perturbative split of the O7-plane \cite{Sen:1997bp}. While such a behavior could not be observed here, the analogy to the stable version of Sen's limit \cite{Clingher:2012rg} gives us at least a hint. There, the elliptic fibre of a Calabi-Yau manifold degenerates into two rational curves in the weak coupling limit, one tracking the open and one tracking the closed string sector. Only in situations where both of these are merged into an elliptic curve is it possible to engineer exceptional singularities. This suggests that exceptional gauge groups not originating purely from the closed string sector are only to be found away from the Kovalev limit, where the Ricci-flat $G_2$ metric is no longer well approximated by the Ricci-flat metric of the acyl Calabi-Yau threefolds and our analysis breaks down. Using the methods of \cite{Pantev:2009de,Braun:2018vhk}, it should be possible to describe such a process at least from a gauge theory perspective. 

TCS $G_2$ manifolds are glued from two acyl Calabi-Yau threefolds times a circle. It is hence tempting to speculate that either one of these circles can be used to reduce an M-Theory compactification to IIA. The existence of a reduction to weakly coupled IIA string theory is equivalent to the existence of a limit in which the $G_2$ manifold in question collapses to a Calabi-Yau threefold. Our results indicate in which cases of TCS $G_2$ manifolds we expect there to be a reduction to IIA, i.e. which TCS $G_2$ manifolds should have a collapsing limit. These are precisely the TCS $G_2$ manifolds we have constructed as lifts of IIA orientifolds, which are those cases where one of the building blocks is (a deformation or resolution) of Voisin-Borcea type.

A closely related observation concerns the case where both building blocks of a TCS $G_2$ manifold are of Voisin-Borcea type. In this case, the $G_2$ manifold $M$ has two different circle collapse limits and there are two associated IIA orientifolds. As these IIA models have identical M-Theory limits, such constructions can be used to engineer a host of new instances of 4D $\mathcal{N}=1$ string-string dualities. Intriguingly, these are such that open and closed string degrees of freedom become interchanged in the duality. It would be interesting to pursue this further. 

As another application, we have shown how to recover different TCS realizations of one and the same $G_2$ manifold by exploiting different K3 fibrations of the type IIA Calabi-Yau orientifold. To the knowledge of the author, these are the first examples with this property and it would be very interesting to describe this phenomenon in more generality and to investigate its implications for other instances of string dualities described in the context of TCS $G_2$ manifolds. The methods for associating a type IIA orientifold to a TCS $G_2$ manifold developed in this work are very similar to the relationship between Spin(7) manifolds constructed from anti-holomorphic involutions of Calabi-Yau fourfolds \cite{joyce1996spin7_new}, and Spin(7) manifolds constructed as `generalized connected sums' in \cite{Braun:2018joh}. Using a similar logic to the one employed here, it must be possible to find examples of Spin(7) manifolds with different inequivalent realizations as a generalized connected sums.

An important continuation of the present work concerns matching the effective actions of the M-Theory reduction and the IIA orientifold reduction. The subsectors of enhanced supersymmetry present in TCS $G_2$ manifolds imply the possibility of setting up a perturbative scheme to describe the $G_2$ effective action from its Calabi-Yau pieces \cite{Guio:2017zfn,Braun:2017csz}. The existence of several TCS realizations lends extra power to such scheme. Furthermore, the present work makes it possible to use the perpendicular approach of construction of the effective action of M-Theory on $G_2$ manifolds from the type IIA orientifold reduction. Starting from a double limit, it appears likely that the combination of both of these approximations provides new insights into the effective action of M-Theory on TCS $G_2$ manifolds.

The effective action of type IIA orientifolds and its M-Theory lift necessarily contains the data of a non-perturvative superpotential. The terms in this superpotential are generated (in the M-Theory language) from membrane instantons wrapped on associative homology spheres in $M$ \cite{Becker:1995kb,Harvey:1999as}. We have ignored such corrections here, but one of the crucial tasks needed for an understanding of the M-Theory lift of IIA orientifolds is to describe and compare the different origins of this superpotential. Superpotentials generated by membrane instantons in the context of M-Theory on TCS $G_2$ manifolds have been studied recently in \cite{Braun:2018fdp,Acharya:2018nbo}.

It remains an open question how to engineer compact $G_2$ manifolds with singularities of codimension seven giving rise to a chiral spectrum of charged matter. Such singularities are absent in TCS $G_2$ manifolds in the limit in which the $G_2$ metric is well approximated by the metrics of the acyl Calabi-Yau threefolds \cite{Guio:2017zfn,Braun:2017uku}, and may only appear pairwise and in such a way that the spectrum is necessarily non-chiral when moving into the interior of the moduli space \cite{Braun:2018vhk}. T-branes in $G_2$ compactifications of M-Theory have recently been proposed as method to engineer chiral spectra in compact models without the need for singularities of codimension seven \cite{Barbosa:2019bgh}. It would be interesting to investigate if our results can be used to explicitly implement these models in TCS $G_2$ manifolds. In a similar vein, it would be interesting to generalize our ideas to non-TCS $G_2$ manifolds and study the M-Theory lift of chiral type IIA orientifolds such as \cite{Cvetic:2001tj,Cvetic:2001nr,Cvetic:2001kk}.

\section*{Acknowledgments}

I wish to thank the many people who asked me about the `M-theory circle' in TCS $G_2$ manifolds. In particular, I would like to thank Cody Long and Jim Halverson for approaching me with many inspiring questions, as well as Bobby Acharya and Roberto Valandro for helpful discussions. The initial parts of this work were performed at the Aspen Center for Physics, which is supported by the National Science Foundation grant PHY-1607611. This work was partially supported by a grant from the Simons Foundation.

\appendix

\section{TCS $G_2$ Manifolds}\label{app:TCSG2}

In this Appendix we review twisted connected sum (TCS) $G_2$ manifolds. They are a special class of $G_2$ manifolds which are glued from pairs of asymptotically cylindrical (acyl) Calabi-Yau threefolds $X_\pm$.  Our aim is mostly to set up notation, see the original literature \cite{MR2024648,Corti:2012kd,MR3109862} or discussions in the physics literature \cite{Halverson:2014tya,Braun:2016igl,Guio:2017zfn} for more details and \cite{Braun:2017uku} for a derivation of the TCS construction from the duality between M-Theory and heterotic strings. 

An acyl Calabi-Yau threefold is a non-compact Calabi-Yau manifold which is diffeomorphic to the product of a K3 surface $S_{0}$ and a cylinder $S^1_b \times I$ outside a compact submanifold. This diffeomorphism must asymptote to an isometry towards the end of the cylinder, see \cite{MR3109862} for details. 

For a compact K\"ahler threefold $Z$ which is fibred by K3 surfaces $S$ from some algebraic family and which satisfies $c_1(Z) = [S]$, an acyl Calabi-Yau threefold $X$ can be constructed by exising a generic fibre
\begin{equation}
 X = Z \setminus S_{0}\, .
\end{equation}
On $Z$ there is a natural restriction map 
\begin{equation}
 \rho: H^{1,1}(Z) \rightarrow  H^{1,1}(S_{0}) \, ,
\end{equation}
which allows us to define
\begin{equation}
\begin{aligned}
N(Z) & := \mbox{im}(\rho) \\
K(Z) & := \ker(\rho)/[S_0] 
\end{aligned}\, .
\end{equation}
We will abbreviate $|N(Z)|=n(Z)$ and $|K(Z)|=k$. We call the algebraic three-folds $Z_\pm$ `building blocks'.

For a pair of acyl Calabi-Yau threefolds $X_\pm$ with cylinder regions $S_{0,\pm} \times S^1_{b,\pm} \times I$, a TCS manifold $M$ is formed by gluing $X_\pm \times S^1_{e, \pm}$ along the cylinderical regions of $X_\pm$ by identifying
\begin{equation}
S^1_{b,\pm} = S^1_{e,\mp}\, ,
\end{equation}
as well as the interval direction and the K3 surfaces $S_{0,\pm}$. The isometry between the K3 surfaces $S_{0,\pm}$ must be such that it implies a hyper-K\"ahler rotation $\varphi$ acting as
\begin{equation}
\begin{aligned}
J(S_{0,\pm}) & = \Re(\Omega^{2,0})(S_{0,\mp}) \\
\Im(\Omega^{2,0})(S_{0,+}) & = -\Im(\Omega^{2,0})(S_{0,-})
\end{aligned} 
\end{equation} 
on the complex structures of $S_{0,\pm}$ inherited from $Z_\pm$.

Under these conditions the resulting topological manifold $M$ admits a Ricci-flat metric with holonomy group $G_2$ which becomes close to the Ricci-flat Calabi-Yau metrics on $X_\pm$ in the limit in which the interval along which the gluing takes place becomes very long (the `Kovalev limit'). We will use the notation
\begin{equation}
 M = \left(Z_- \times S^1_{e-} \right) \#_\varphi \left(Z_+ \times S^1_{e+} \right) \, .
\end{equation}
as a short-hand for this construction. 

The integral cohomology groups of $M$ can be expressed as \cite{Corti:2012kd}
\begin{equation}\label{eq:bettitcsz}
\begin{aligned}
H^1(M,\mathbb{Z}) & =   0 \\
H^2(M,\mathbb{Z}) & =  N_+ \cap N_- \oplus K(Z_+) \oplus K(Z_-) \\
H^3(M,\mathbb{Z}) & = \mathbb{Z}[S] \oplus \Gamma^{3,19} /(N_+ + N_-) \oplus (N_- \cap T_+) \oplus (N_+ \cap T_-)\\
& \hspace{1cm} \oplus H^3(Z_+)\oplus H^3(Z_-) \oplus K(Z_+) \oplus K(Z_-) \\
H^4(M,\mathbb{Z}) & = H^4(S) \oplus (T_+ \cap T_-) \oplus \ \Gamma^{3,19} /(N_- + T_+) \oplus \Gamma^{3,19} /(N_+ + T_-) \\
& \hspace{1cm} \oplus  H^3(Z_+)\oplus H^3(Z_-) \oplus K(Z_+)^* \oplus K(Z_-)^* \\
H^5(M,\mathbb{Z}) & = \Gamma^{3,19} /(T_+ + T_-) \oplus K(Z_+) \oplus K(Z_-) \,.
\end{aligned}
\end{equation}
Here $T = N^\perp$ in $H^2(K3,\mathbb{Z})$, $N_\pm = N(Z_\pm)$ and $T_\pm = T(Z_\pm)$. Under the condition that 
\begin{equation}
N_\pm \otimes \mathbb{R} = \left(N_\pm\otimes \mathbb{R}  \cap N_\mp\otimes \mathbb{R}  \right) \oplus \left(N_\pm\otimes \mathbb{R}  \cap T_\mp\otimes \mathbb{R} \right)
\end{equation}
 (`orthogonal gluing') the simplified relation
\begin{equation}\label{eq:bettitcs}
b^2 +b^3 = 23 + 2 \left[ k(Z_+) + k(Z_-) + h^{2,1}(Z_+) + h^{2,1}(Z_-) \right]
\end{equation}
for the sum of the Betti numbers holds.

\section{Acyl Calabi-Yau Manifolds, Tops, and \\ Anti-holomorphic Involutions}\label{app:buildingblocks}

In this section, we will detail the construction of building blocks $Z$ with particular emphasis on the geometries occuring in M-Theory lifts of IIA orientifolds. 

Building blocks may be realized from blowups of semi-Fano threefolds \cite{MR3109862}. Here, one blows up along the intersection of two anti-canonical hypersurfaces in the semi-Fano threefold $F$. Concretely, this results in a threefold $Z$ which is described as a hypersurface
\begin{equation}\label{eq:cortibuildingblocks}
z_1 P =  Q z_2
\end{equation}
in $\mathbb{P}^1 \times F$. Here $P,Q$ are sections of $-K_F$, so that $c_1(Z) = [z_1]$. By projecting the ambient space $\mathbb{P}^1 \times F$ to $\P^1$, $Z$ carries the structure of a K3 fibration. The fibres are from the algebraic family of K3 hypersurfaces in $F$.

The above description has a natural realization in terms of toric geometry, which has been described in \cite{Braun:2016igl}. If $F$ is a toric variety with a fan $\Sigma$ which can be obtained by triangulating a reflexive polytope $\Delta^\circ_F$, the threefold \eqref{eq:cortibuildingblocks} is found in complete analogy to \cite{Batyrev:1994hm} by starting from a four-dimensional polytope $\Diamond$, which is given as the Minkowski sum 
\begin{equation}
\Diamond = \Delta_F + (0,0,0,-1) \, .
\end{equation}
Here, $\Delta_F$ is the polar dual of $\Delta^\circ_F$. A refinement of the normal fan of $\Diamond$ then yields $F \times \P^1$, and $\Diamond$ becomes the Newton polytope of the hypersurface equation \eqref{eq:cortibuildingblocks}. 

The above observation has a natural generalization to `projecting tops', which are pairs of polytopes $\Diamond^\circ$ and $\Diamond$ obeying 
\begin{equation}\label{eq:topsduality}
 \begin{aligned}
 & \langle \Diamond, \Diamond^\circ \rangle \geq -1 & \\
  \langle  \Diamond,\nu_e \rangle \geq 0 \hspace{.5cm} & &  \langle m_e, \Diamond^\circ\rangle \geq 0
\end{aligned}\,.
\end{equation}
with the choice $m_e = (0,0,0,1)$ and $\nu_e = (0,0,0,-1)$. Furthermore, projecting $\Diamond$ and  $\Diamond^\circ$ to the first three coordinates must produce a pair of reflexive polytopes $\Delta_F$ and $\Delta_F^\circ$ which are equal to the polytopes found when intersecting $\Diamond^\circ$ and $\Diamond$ with the hyperplanes perpendicular to $m_e$ and $\nu_e$.

The polytope $\Diamond$ defines a compact but generally singular toric variety through its normal fan $\Sigma_n(\Diamond)$, together with a hypersurface $Z_s(\Diamond)$ (actually, $\Diamond$ defines a family of hypersurfaces and $Z_s(\Diamond)$ denotes a generic member of this family). The variety $Z_s(\Diamond)$ can be crepantly resolved into a smooth manifold $Z(\Diamond)$ by refining the fan $\Sigma \rightarrow \Sigma_n$ using all of the lattice points on $\Diamond^\circ$ as rays. As shown in \cite{Braun:2016igl}, such manifolds have all of the properties of building blocks. Concretely, the defining equation of the resolved hypersurface is
\begin{equation}\label{HypSurf}
Z(\Diamond): \,\,\, 0 = \sum_{m \in \Diamond} c_m z_0^{\langle m, \nu_0 \rangle}\prod_{\nu_i \in \Diamond^\circ} z_i^{\langle m, \nu_i\rangle +1} \, .
\end{equation}
Here, $m$ runs over all of the lattice points on $\Diamond$ and $c_m$ are generic complex coefficients. The $z_i$ are homogeneous coordinates associated with lattice points $\nu_i$ on $\Diamond^\circ$, and $z_0$ is the homogeneous coordinate associated with the ray through $\nu_0 = (0,0,0,-1)$. The first Chern class of  
$Z(\Diamond)$ is equal to the class $[z_0]$ and defines a K3 surface with trivial normal bundle. The Hodge numbers of $Z(\Diamond)$, as well the ranks of the lattices $N$ and $K$ can be computed in purely combinatorial terms, details can be found in \cite{Braun:2016igl,Braun:2017ryx}.

The definition of projecting tops \eqref{eq:topsduality} implies that for any $\Diamond_a$ and $\Diamond_b$ for which $\Delta_{a,F} = \Delta_{b,F}$, there is an associated reflexive polytope $\Delta$ \cite{Candelas:2012uu,Braun:2017ryx}. The reflexive polytope $\Delta$ is given by the union of $\Diamond_a$ and $\bar{\Diamond}_b$, where $\bar{\Diamond}$ is the same as the polytope $\Diamond$ with the fourth coordinate inverted. Furthermore, the polar dual $\Delta^\circ$ of $\Delta$ is given by the union of $\Diamond_a^\circ$ and $\bar{\Diamond}_b^\circ$. 

The geometrical meaning of these relations is simple: $\Delta^\circ$ and $\Delta$ describe a K3-fibred Calabi-Yau threefold $X(\Delta)$ which has a degeneration limit in which it decomposes into two building blocks $Z(\Diamond_a)$ and $Z(\Diamond_b)$ \cite{Braun:2017ryx}. We may also think of cutting $X(\Delta)$ into two halves $X_a$ and $X_b$ by cutting the base of the K3 fibration of $X(\Delta)$ into two halves. For a suitable choice this can be done such that 
\begin{equation}
\begin{aligned}
X_a & = Z(\Diamond_a) \setminus S_{0,a} \\
X_b & = Z(\Diamond_b) \setminus S_{0,b} 
\end{aligned}\,\, ,
\end{equation}
i.e. we can think of $X(\Delta)$ as being glued from two acyl Calabi-Yau threefolds. The data of the K3 fibration on $X(\Delta)$ is separated into two pieces which are captured by $Z(\Diamond_a)$ and $Z(\Diamond_a)$. We will use the notation
\begin{equation}
X(\Delta) =  Z(\Diamond_a)\, \#\, Z(\Diamond_b) \, 
\end{equation}
in this situation.

In case $\Diamond_a = \Diamond_b$, $X(\Delta)$ allows an anti-holomorphic involution of the type considered in Section \ref{sect:ahinvonk3fib}. The homogeneous coordinates $[b_1:b_2]$ on the base $\P^1$ of $X(\Delta)$ are \cite{Avram:1996pj,Kreuzer:2000qv}:
\begin{equation}
[b_1:b_2] = [\prod_{\nu_i \in \Diamond_a^\circ} z_i^{\langle m_e,\nu_i \rangle}: \prod_{\nu_i \in \bar{\Diamond}_a^\circ} z_i^{-\langle m_e,\nu_i \rangle}] \, .
\end{equation}
Every coordinate $\nu$ on $\Diamond^\circ_a$ with associated homogeneous coordinate $z(\nu)$ has a counterpart $R \nu$ on $\bar{\Diamond}_a^\circ$ with homogenous coordinate $z(R\nu)$. Here, $R$ is the matrix $\mbox{diag}(1,1,1-1)$. Hence we can map 
\begin{equation}
\begin{aligned}
b_1 \rightarrow \bar{b}_2 \\
b_2 \rightarrow \bar{b}_1  
\end{aligned}
\end{equation}
by mapping 
\begin{equation}
z(\nu) \rightarrow \overline{z(R\nu)} 
\end{equation}
for all $\nu$ not contained in $\Delta^\circ_F$. The action on $z(\nu)$ for $\nu \in \Delta^\circ_F$ can be freely chosen (up to the requirement of being an isometry of the ambient space of the hypersurface). After redefining coordinates to
\begin{equation}
\begin{aligned}
b_1' &= b_1 + b_2 \\ 
b_2' &=b_1 - b_2
\end{aligned}
\end{equation}
this precisely captures the class of involutions used in Section \ref{sect:ahinvonk3fib}. We hence recover the fact that projecting tops are associated with such involutions, and it hence comes as no surprise that the building block $Z(\Diamond)$ features in the $G_2$ lift of such orientifolds.

\section{Nikulin Involutions and Voisin-Borcea Threefolds}\label{app:nikulin}

An isometry $\sigma$ of a K3 surface $W$ is called a non-symplectic involution if its induced action on the K\"ahler form $J$ and holomorphic two-form $\Omega^{2,0}$ is 
\begin{equation}
\sigma^*:\,\,\,
\begin{aligned}
&J &\rightarrow& \,\,\,\,\,\,J \\
&\Omega^{2,0} &\rightarrow& \,\,\,\,\, -\Omega^{2,0} \\
\end{aligned} 
\end{equation}
We may decompose the middle cohomology of $W$ into even and odd eigenspaces, which defines the lattices
\begin{equation}
H^2(W,\mathbb{Z}) \supseteq  H^2_+(W,\mathbb{Z}) \oplus H^2_-(W,\mathbb{Z})
\end{equation}
Let $S = H^2_+(W,\mathbb{Z})$ and $r=\rk(S)$. Then 
\begin{equation}
S^*/S = \mathbb{Z}_2^a \, . 
\end{equation}
The inner form on $H^2(W,\mathbb{Z})=U^{\oplus 3} \oplus (-E_8)^{\oplus 2}$ furthermore determines the discriminant form
\begin{equation}
q: S^*/S \rightarrow \mathbb{Q}/2\mathbb{Z} \, , 
\end{equation}
and we set $\delta=0$ if it is even and $\delta=1$ otherwise. 

The triple of numbers $(r,a,\delta)$ is sufficient to characterize the involution $\sigma$ \cite{nikulin1976finite,0025-5726-14-1-A06,2004math......6536A}, and the fixed locus $L$ of $\sigma_{r,a,\delta}$ is given as follows
\begin{itemize}
 \item if $(r,a,\delta) = (10,10,0)$, $L$ is empty (this is the Enriques involution)
 \item if $(r,a,\delta) = (10,8,0)$, $L$ consists of two elliptic curves
 \item for all other cases, $L$ consists of a Riemann surface of genus $g$ and $f-1$ rational curves, where
\begin{equation}
\begin{aligned}
g &=  (20-r-a)/2 + 1\\
f &= (r-a)/2  +1   
\end{aligned}
\end{equation}
\end{itemize}

For every single one of the non-simplectic involutions $\sigma_{r,a,\delta}$ introduced above, there exists an associated Calabi-Yau threefold, which is contructed by resolving the orbifold
\begin{equation}
(W \times E)/(\sigma, -) 
\end{equation}
 for a K3 surface $W$ and an elliptic curve $E$. The fixed locus of $(\sigma, -)$ on $W \times E$ consists of four copies of $L$, and it was shown in \cite{voisin1993miroirs,borcea1996k3} that there always is a resolution $Y_{r,a,\delta}$ such that
\begin{equation}
\begin{aligned}
 h^{1,1}(Y_{r,a,\delta}) & = 1 + r + 4f \\
 h^{2,1}(Y_{r,a,\delta}) & = 1 + 4g + 20-r
\end{aligned} \, .
\end{equation}
In particular, the Calabi-Yau threefolds $Y_{r,a,\delta}$ come in mirror pairs,
\begin{equation}
Y_{r,a,\delta}^\vee = Y_{20-r,a,\delta} \, .
\end{equation}

\bibliographystyle{plain}

\end{document}